\documentclass[12pt,preprint]{aastex}
\usepackage{epsf}

\shorttitle{Cosmic Ray Modified Shocks}
\shortauthors{Kang \& Jones}

\def\etal{{\it et al.~}}
\def\eg{{\it e.g.,}}
\def\ie{{\it i.e.,~}}
\def\kms{~{\rm km~s^{-1}}}
\def\cm3{~{\rm cm^{-3}}}

\def\lsim{\mathrel{  
        \raise0.3ex\hbox{$<$}\kern-0.75em{\lower0.65ex\hbox{$\sim$}}}}
\def\gsim{\mathrel{
        \raise0.3ex\hbox{$>$}\kern-0.75em{\lower0.65ex\hbox{$\sim$}}}}

\begin{document}
\title{Efficiency of Nonlinear Particle Acceleration at Cosmic Structure Shocks}

\author{Hyesung Kang}
\affil{Department of Earth Sciences, Pusan National University,
    Pusan 609-735, Korea} 
\email{kang@uju.es.pusan.ac.kr}

\and

\author{T.W. Jones}
\affil{Department of Astronomy, University of Minnesota, Minneapolis, 
      MN 55455}
\email{twj@msi.umn.edu}

\altaffiltext{1}{Submitted to the Astrophysical Journal}

\begin{abstract}
We have calculated the evolution of cosmic ray (CR) modified astrophysical 
shocks for a wide range of shock Mach numbers and shock speeds 
through numerical simulations of 
diffusive shock acceleration (DSA) in 1D quasi-parallel plane shocks.
The simulations include thermal leakage injection of seed CRs, as well as
pre-existing, upstream CR populations. Bohm-like diffusion is assumed.
We model shocks similar to those expected around
cosmic structure pancakes as well as other accretion shocks driven by
flows with upstream gas temperatures in the range $T_0=10^4-10^{7.6}$K 
and shock Mach numbers spanning $M_s=2.4-133$.
We show that CR modified shocks evolve to time-asymptotic states by the time 
injected particles are accelerated to moderately relativistic energies 
($p/mc \gsim 1$),
and that two shocks with the same Mach number, but with different shock
speeds, evolve qualitatively similarly when the results are presented in
terms of a characteristic diffusion length and diffusion time.
We determine and compare the ``efficiencies'' of CR acceleration in our
simulated shocks by calculating the ratio of the total CR energy generated 
at the shock to the total kinetic energy that would pass through the shock 
over time in its initial frame of reference.
For these models the time asymptotic value for this efficiency
ratio is controlled mainly by shock Mach number, 
as expected from the aforementioned similarity in CR shocks. 
In the presence of a pre-existing CR population, shock evolution proceeds 
similarly to that for higher thermal injection rates compared to thermal 
leakage CR sources alone. This added contribution has little or no
impact on the postshock or CR properties of the high 
Mach number shocks simulated.
The modeled high Mach number shocks all evolve towards efficiencies
$\sim 50$\%, regardless of the upstream CR pressure.
On the other hand, the upstream CR pressure increases 
the overall CR energy in moderate strength shocks ($M_s \sim {\rm a~few}$),
since it is a significant fraction of the shock ram pressure. 
These shocks have been shown to dominate dissipation during cosmic 
structure formation, so such enhanced efficiency 
could significantly increase their potential importance as sources of cosmic rays.

\end{abstract}


\section{Introduction}

Collisionless shocks form ubiquitously in tenuous cosmic plasmas via
collective, electromagnetic viscosities. The formation process of such shocks 
inevitably produces suprathermal particles in addition to thermal particles 
with Maxwellian velocity distributions \citep{blaeic87, Jones91, Draine93}. 
These nonthermal particles can be further accelerated to very high energies
through the interactions with
resonantly scattering Alfv\'en waves in the converging flows across a shock,
\ie by diffusive shock acceleration (DSA) \citep{dru83,blaeic87,maldru01}.
Detailed nonlinear treatments of DSA predict that
a small fraction of incoming thermal particles can be injected
into the CR population, and that a significant fraction of the shock kinetic energy 
can be transferred to CRs \citep[\eg][]{ebj95, kangetal02}.

Simulations of DSA in spherical supernova remnants (SNRs) indicate that CRs
can absorb up to 50\% of the initial blast energies \citep[\eg][]{berzvolk97, 
berzvolk00}.  Support for rapid and efficient DSA in that
setting has been provided by recent X-ray 
observations of young SNRs such as SN1006 and Cas A that indicate 
the presence of short-lived, TeV electrons emitting nonthermal synchrotron emission 
immediately inside the outer SNR shock \citep{koyama95, allen97, bamba03}. 
On a galactic scale it is well known that the CR energy density is 
comparable to the gas 
thermal energy density in the interstellar medium and plays important 
dynamical roles in the evolution of our Galaxy. 
Although the Galactic CRs are commonly believed to be accelerated mostly
at SNR shocks, CR acceleration is probably important 
in all shock-heated astrophysical plasmas. 

CR populations are also indicated by extended, diffuse nonthermal
emissions in some galaxy clusters \citep[\eg][and references therein]{min01}.
One likely contribution to that CR population is large scale
shocks associated with the formation of the clusters. 
According to hydrodynamic simulations of large scale
structure formation in the universe,
nonlinear structures such as pancakes, filaments, and knots
are surrounded by ``external'' accretion shocks and contain 
complex webs of ``internal'' flow shocks including, but not limited to, 
accretion shocks around 
individual clusters and merger shocks inside clusters \citep{min00}.
Recently, \citet{ryu03} studied the characteristics of such cosmic shocks
and showed that they have a wide range of physical
parameters with shock speeds, $u_s$, up to $\sim 3000\kms$,
preshock temperatures of $10^4< T_0< 10^8$ K, and
Mach numbers up to a few 100.
They showed that in the present universe
the mass inside nonlinear structures has been
shocked approximately twice on average over cosmic time, and that 
shocks with $2 \la M \la 4$
have contributed $\sim 1/2$ of the total energy dissipated at the shocks.
Utilizing nonlinear DSA model calculations of \citet{kangetal02}, Ryu \etal estimated
that the ratio of CR energy to gas thermal energy 
resulting from dissipation at cosmological shocks is $\sim1/2$. 
Hence, the CR energy density could be dynamically important 
in the intracluster medium, just as in the interstellar medium of our Galaxy
and might have had some dynamical influences on large scale structure 
formation.

The above ``cosmic structure shocks'' are an inevitable consequence of
gravitational collapse, and are central to virialization
of diffuse baryonic matter. The DSA process behind our
present discussion depends, of course, on the existence of a 
magnetic field in the shock vicinity to mediate Alfv\'enic 
turbulence that scatters the CRs. The effectiveness of the scattering
at a particular particle momentum, $p=\gamma \upsilon m$, can be expressed in terms of 
the DSA diffusion time, $t_d(p) = \kappa(p)/u^2_s$,
where $\kappa = (1/3) r_g \upsilon  (B/\delta B)^2$ is the spatial
diffusion coefficient, $r_g = pc/(eB)$ is the particle gyroradius,
$B$ is the magnetic field, and $\delta B$ characterizes the
magnitude of resonant wave magnetic fields.
Particles will generally be accelerated from low energies
to $E \sim pc$ on timescales a few times $t_d(p)$ (see equation \ref{tacc}).
Efficient transfer of kinetic energy to CRs typically becomes
apparent in DSA simulations by the time transrelativistic CRs are accelerated,
establishing the relevant time scale for the introduction of
nonlinear feedback into the shock dynamics.

Evidence is strong that magnetic fields exist in clusters 
at levels exceeding at least a few $\times 10^{-7}$Gauss \citep[\eg][]{kron01}. 
Fields of that strength lead, with Bohm diffusion ($\delta B = B$),
to diffusion times for transrelativistic CRs, 
$t_d = (1/3) pc\upsilon/(eB u_s^2)$
of the order of a few years for $p \sim mc$ in fast cosmic shocks with
speeds $\sim 10^3\kms$. 

Direct evidence for microGauss-level magnetic
fields where most cosmological shocks form, that is, outside individual
clusters, is currently quite limited. 
However, there are good reasons to expect fields
adequate for efficient DSA in most cosmic shocks. For example, 
even in the absence of any primordial seed field, magnetic fields
are likely to be spontaneously generated 
at ``external cosmic shocks''
by some combination of the Biermann battery \citep{kuls97,rkb98}
and the Weibel instability \citep{sch03}. Whether such
seed fields are initially strong enough to reduce
diffusion times to cosmologically interesting values or not,
quasi-linear plasma theory and simulations show that
the streaming motion of suprathermal particles 
can induce resonant Alfv\'en wave generation and 
strong field amplification upstream of collisionless shocks
\citep[\eg][]{Bell78,Quest88, lucek00, bell01}.
\cite{bell01} have argued
that magnetic fields can ultimately be amplified to produce magnetic
pressures that are a substantial fraction of the ram pressure through
the shock, $\rho_0 u^2_s$. That limit would correspond to field strengths
$\sim 4\times 10^{-7} h (1+\delta)^{1/2} u_{s,3}(1+z)^{3/2}$Gauss, where $h$ is
the Hubble parameter in units of 100 km/sec/Mpc, 
$\delta=\delta\rho/\rho$ is the local overdensity ratio, 
$z$ is the redshift of the epoch, and $u_{s,3}$ is 
the shock speed in units of $10^3\kms$.
A baryon density fraction, $\Omega_b=0.04$, is assumed. 
Even magnetic fields only one percent of this value would be sufficient
to accelerate CRs to relativistic energies, and, hence establish
nonlinear CR feedback on timescales of thousands
of years in the shocks of interest. 

Once shocks develop nonlinear properties in response to CR
feedback, there are several important characteristics that distinguish them
from more familiar gasdynamic shocks:
1) CR shocks continue to evolve over relatively long times
and broaden as they do so. 
While full thermalization takes place instantaneously at a simple, 
discontinuous jump in an ``ideal'' gasdynamic shock, CR acceleration
and the corresponding modifications to the underlying flow 
depend on suprathermal particles passing back and forth diffusively 
across the shock structure. 
These processes develop, therefore, on the 
diffusion time scale, $t_d$, and diffusion
length scale, $x_d = t_d u_s$, as alluded to earlier.
Generally, these scales are expected to be increasing functions of
particle momentum, so CR acceleration and
shock evolution rates slow over time.
For very high energy particles DSA time and length scales often approach
the age and the radius of curvature of astrophysical shocks.
2) CR diffusion upstream of the
shock discontinuity leads to strong pressure gradients
in a shock precursor, enhancing the total compression
through the shock transition over that in the ``viscous'' subshock.
The total compression through a high-Mach-number CR shock can
greatly exceed the canonical value $r=(\gamma_g+1)/(\gamma_g-1)$ 
for strong gasdynamical shocks \cite[\eg][]{berzell99, maldv00} 
(where $\gamma_g$ is the gas adiabatic index).
3) Both the effective compressibility of the combined thermal-CR plasmas and 
the mean CR diffusion
coefficient increase over time as relativistic CRs absorb more energy
and as escaping CRs remove energy from the shock structure.
Details of these properties depend on the CR momentum distribution.
Thus, downstream states of the CR modified shock cannot be found from
simple ``shock jump conditions", but rather have to be integrated 
time-dependently from given initial states or, for steady solutions,
found in terms of some predefined limits to the CR spectrum 
(\eg an upper momentum cutoff).
4) Energy transfer to the CRs rather
than to the thermalized gas in the shock transition reduces
the downstream thermal energy of a CR modified shock
compared to a gasdynamic shock with the same shock speed and Mach number. 
The Mach number of the gas subshock is also reduced. Even in 
a shock with a very large total Mach number, the subshock Mach number
and compression ratio are typically $\sim 3$ \citep[\eg][]{mal98,berzell99,kangetal02}.

In order to study these processes closely
we recently developed a numerical scheme that self-consistently 
incorporates a thermal leakage CR injection model based on the analytic, 
nonlinear calculations of \cite{mal98} \citep{gies00}
and implemented it into a combined Adaptive Mesh Refinement (AMR)
gas dynamics
and CR diffusion-convection code \citep{kangetal02}. As described
in the following section, there is only one rather well constrained
parameter needed to express this injection model; namely, the
strength of nonlinear MHD waves immediately downstream of the
plasma subshock with respect to the upstream longitudinal magnetic
field strength.

Previously, we applied our new code to a preliminary investigation of
cosmic shocks emerging in large scale structure formation of the universe
\citep{kang02, kang03}. 
Utilizing the Bohm diffusion model in 1D quasi-parallel, plane-parallel shocks
we showed that the resulting CR modified shocks evolve roughly ``self-similarly''
if the shock structure is expressed in terms of diffusion time and length scales,
and that the time asymptotic states depend mainly on the
shock Mach number, with only a weak dependence on the parameter needed
in the injection model.
We found for strong shocks
that about $10^{-3}$ of incoming thermal particles are injected 
into the CR population with our thermal leakage injection model
over the long term,
and up to $\sim 1/2$ of the shock kinetic energy flux measured
in the initial shock frame is transferred to CRs. 
In the present contribution we confirm and extend our previous studies 
by including simulations of shock models with a broader range of 
physical parameters suitable for application to cosmic structure
formation shocks, and also by including the influence of a pre-existing,
upstream CR population, since many of the shocks of interest will
probably form in previously shocked plasma.
Our numerical simulations here are limited to quasi-parallel shocks
in which the mean magnetic field lines are parallel to the direction of
shock propagation. In quasi-planar shocks this is not a strong limitation
so long as the obliquity of the magnetic field
is small enough to keep its intersection subluminal and so long
as the magnetic field is too weak to be dynamically important
\citep[\eg][]{jok82,dru83}. 
In addition, since
streaming motions inside cosmic sheets and filaments appear to stretch the field lines
\citep[\eg][]{rkb98}, many structure formation ``internal'' shocks are
likely to be quasi-parallel. The principle concern may be the influence of
obliquity on the injection of low energy CRs at the shocks, as we discuss
below.

In the following section 
we briefly outline our numerical methods in the CR/AMR hydrodynamics code. 
The simulation results are presented and discussed in \S 3,
followed by a summary in \S 4.


\section{Numerical Calculations}

\subsection{The CRASH code}
Our conservative, Eulerian CR/AMR hydrodynamics code, CRASH (Cosmic-Ray Amr SHock), 
solves the gasdynamic equations with CR pressure terms added
for one dimensional
plane-parallel geometry,
along with the diffusion-convection equation
for the CR momentum distribution function, $g(p) = f(p)p^4$.
The full CR shock transition includes a very wide range of length scales associated
with the particle diffusion lengths that must be resolved to follow
the shock evolution properly. To include CRs injected by
thermal leakage, these scales must extend down close
to the subshock thickness.
In order to handle this dynamic range efficiently,
an adaptive mesh refinement (AMR) technique is applied to regions
around gasdynamic shocks,
which are tracked as discontinuous jumps by a front-tracking
method \citep{kang01}.
An additional equation for the ``Modified Entropy'',
$S= P_g / \rho^{\gamma_g-1}$, is 
solved to follow accurately the adiabatic changes outside of shocks, 
particularly in the precursor region  of strong shocks \citep{kangetal02}.
Readers are referred to these previous papers for further numerical
details.

\subsection{The Bohm Diffusion Model}
DSA, along with the time and length scales for CR acceleration,
and subsequent shock modification, depend on the spatial
diffusion coefficient for the CRs.
The diffusion coefficient can be expressed
in terms of a mean scattering length, $\lambda$, as
$\kappa(x,p) = {1\over 3} \lambda \upsilon$, where $\upsilon$
is the particle velocity.
As noted earlier, for Alfv\'en wave scattering one can write
$\lambda \approx r_g (B/\delta B)^2$ 
\cite[\eg][]{ski75}.
The Bohm diffusion model, representing a saturated wave spectrum
($\delta B \sim B$) and the  minimum diffusion coefficient,
gives $\kappa_{\rm B} = (1/3) r_{\rm g} \upsilon$.
If we assume ``Bohm-like'' diffusion with
$\lambda = \zeta \times r_g$ with $\zeta \ge 1$, the diffusion coefficient
can be expressed as
\begin{equation}
\kappa(\rho,p) = \kappa_{\rm n}\left(\frac{\rho_0}{\rho}\right) 
{ p^2\over (p^2+1)^{1/2} },
\label{kapeq}
\end{equation}
where hereafter momentum, $p$, is expressed in units of $mc$ and
$\kappa_{\rm n} = \zeta \times (1/3) m c^2/(eB)$, 
represents the diffusion coefficient far upstream of the shock for CRs with 
$p \approx 1.3$. 
The assumed density dependence for $\kappa$ accounts for compression of the 
perpendicular component of the wave magnetic field and also
inhibits the acoustic instability in the precursor of highly modified 
CR shocks \citep{drufal86,kanjonryu92}.  
Also, hereafter, we use the subscripts '0', '1', and '2' to denote
conditions far upstream of the shock, immediately upstream of the
gas subshock and immediately downstream of the subshock, respectively.
Thus, $\rho_0$ represents the far-upstream gas density.

The so-called DSA diffusion time, $t_d = \kappa/u_s^2$, defined in the
introduction represents the time scale for hydrodynamical advection across
the CR diffusion length, $x_d$, which, in turn represents the length on
which CR diffusion upstream is balanced by advection towards the subshock.
Conveniently, $t_d$ also provides a natural time unit to measure the
acceleration time for individual CRs. The residence time for individual
particles on each side of the shock is proportional to $x_d/\upsilon$, while
the CR fractional momentum  gain per shock crossing pair is proportional to
$\Delta u/\upsilon$. The resulting acceleration time scale for a particle to
reach momentum $p$ \citep[\eg][]{dru83} is defined as 
\begin{equation}
\tau_{acc}(p) 
= {3\over {u_1-u_2}} ({\kappa_1\over u_1} + {\kappa_2\over u_2})
\approx {8M_s^2\over {M_s^2-1}} t_d(p)~ .
\label{tacc}
\end{equation}
The approximate, Mach-number-based expression in equation (\ref{tacc})
is based on the compression
through a gasdynamic shock for $\gamma_g=5/3$ and sets $u_s = u_1$.
Once again, $t_d(p) = \kappa(p)/u_s^2$.
Thus, the CR acceleration time depends on both the speed and the Mach number
(actually compression) of the shock in addition to the diffusion coefficient.
In the limit of strong shocks ($M_s\gg1$), however, it becomes independent
of the shock Mach number, and is about an order of magnitude
greater than the nominal diffusion time, \ie $\tau_{acc}\approx 8 t_d$. 
We note, however, that in CR modified shocks
the compression factor across the total shock structure can be greater 
than that of strong gasdynamic shocks, and that higher energy CRs 
with greater diffusion length see a greater compression ratio.
Consequently, the mean acceleration timescale  at a fixed
CR energy is somewhat smaller than that given 
in equation (\ref{tacc}).  

\subsection{The Thermal Leakage Injection Model}

In the ``thermal leakage'' model for CR injection at shocks, most of 
the downstream thermal
protons are locally confined by nonlinear MHD waves and only particles
well into the tail of the postshock Maxwellian distribution can
leak upstream across the subshock
\citep[\eg][]{elleich85, mal98, maldru01}.
In particular, ``leaking'' particles not only must have velocities large
enough to swim against the 
downstream flow in order to return across the shock, they must avoid being 
scattered by
the MHD waves that mediate the plasma subshock. The latter effect
further strongly
filters particles against leaking. Thus, the ratio of the breadth of the 
postshock thermal velocity distribution to the
downstream flow velocity in the subshock rest-frame is central
to the injection problem. \citet{malvol98} and 
\citep{mal98} developed a nonlinear analytic
model to incorporate these features, and we have adapted that model
to our simulations.
In order to model this injection process numerically \citet{gies00}
constructed a ``transparency function'',
$\tau_{\rm esc}(\epsilon, \upsilon)$ 
that expresses the probability of supra-thermal
particles at a given velocity, $\upsilon$, leaking upstream through the postshock
MHD waves. 
One free parameter, defined  by \citet{malvol98},
controls this function; namely, $\epsilon = B_0/B_{\perp}$, 
which is the ratio of the amplitude
of the postshock MHD wave turbulence $B_{\perp}$ to the general magnetic field
aligned with the shock normal, $B_0$.
As noted by \citet{malvol98} this parameter is rather well constrained, 
since $0.3\lsim \epsilon \lsim 0.4$
is indicated for strong parallel shocks.
However, such large values of $\epsilon$ lead to very 
efficient initial injection and
most of the shock energy is quickly transferred to the CR component for strong shocks of
$M_s \gsim 30$ \citep{kang02}, causing a numerical problem at the very
early stage of simulations.
Consequently, we reduced the shock transparency by setting
$\epsilon=0.2$ in this study. 
\citet{kang03} showed that time asymptotic states of the CR shocks
depend only weakly on the values of $\epsilon$ in this range. 
On the other hand, smaller $\epsilon$ values may also be more in 
tune with behaviors in moderately oblique shocks, since
finite magnetic field obliquity also reduces shock transparency
and the rate of thermal injection \citep[\eg][]{ebj95,vbk03}.
We note that \citet{vbk03} estimate that the injection rate for angles $\gsim 35^\circ$
falls below the critical injection rate for efficient
shock modification, $\xi \sim 6\times10^{-5}$ (as defined in equation \ref{xieq}), and CR
acceleration becomes very inefficient.

\subsection{Normalization of Physical Variables}
The ideal gasdynamic equations 
in 1D planar geometry 
do not contain any intrinsic
time and length scales, but in CR modified shocks the CR acceleration and
the precursor growth can be characterized by the diffusion scales,
$t_d(p)$ and $x_d(p)$.
$\kappa_n=\kappa(p\approx1.3)$ provides a useful canonical value, 
since nonlinear feedback from CRs to the underlying flow becomes
significant typically by the time transrelativistic CRs are accelerated.
Consequently, we normalize our shock evolution times and structure scales
by $t_{\rm n}=\kappa_{\rm n}/u_{\rm n}^2$, and 
$x_{\rm n}=\kappa_{\rm n}/u_{\rm n}$, respectively,
where $u_{\rm n}$ is a characteristic flow speed. 
One expects intrinsic similarities in the dynamic evolution and structure
of two CR shock models with the same Mach number, but with different shock speeds,
{\it so long as the results are expressed in terms of $t_{\rm n}$ and
$x_{\rm n}$.}
This approach removes the need for any particular choice
in $\kappa_{\rm n}$, but we note for convenience
that $\kappa_{\rm n} = 3.1\times 10^{23}~\zeta/B_{-7}~{\rm cm~ s^2}$,
where $B = B_{-7}\times 10^{-7} {\rm Gauss}$. Then for example, a
characteristic diffusion time, $t_{\rm n} = \kappa_{\rm n}/u^2_{\rm n}
=3.1\times 10^{7}~\zeta/(B_{-7} u^2_{{\rm n},3})$ sec, 
and $x_{\rm n} = 3.1\times 10^{15}~\zeta/(B_{-7} u_{{\rm n},3})$ cm, where 
$u_{\rm n} = u_{{\rm n},3}\times 10^3 $ km/sec. 

In the simulations described below we express pressures and 
energy densities in units $P_{\rm n}=\rho_{\rm n} u_{\rm n}^2$,
where $\rho_n$ will be an appropriate fiducial gas density. 
Again, for reference, we can express these quantities in practical
units in terms of the preshock cosmic overdensity ratio, $\delta$, as
$\rho_{\rm n} \approx  7.5\times 10^{-31} (1+\delta) h^2 (1+z)^3 {\rm g cm^{-3}}$ 
and $P_{\rm n} \approx 7.5\times 10^{-15} h^2 (1+\delta) (1+z)^3 u^2_{n,3} 
{\rm dyne~ cm^{-2}}$, with the same cosmological choices mentioned in
the introduction.

When the shock speeds are less than a percent or so of the speed of
light, freshly injected CRs are essentially nonrelativistic
($p_{\rm inj} \propto \sqrt{T_2} \propto (u_s/c)$), so
both the diffusion coefficient and CR partial pressure
take power law forms; thus, allowing self-similar behaviors in
modified shock evolution.
On the other hand, 
once the shock speed exceeds a percent or so of the speed of light,
the full expression for the particle speed, $\upsilon = p/\sqrt{p^2+1}$,
must be applied, so
neither the diffusion coefficient nor the
CR partial pressure permit self-similar scalings.
Consequently, the physical shock speed does influence
the evolution of the fastest shocks we simulated in this study.

\subsection{Simulation Set Up and Model Parameters }
We model shocks that form in an accretion flow entering through the 
right boundary in a 1D simulation box,$[0,x_{\rm max}]$. The inflow
has constant density, speed, and pressure. 
This flow reflects off the left boundary, or piston, to
initiate the shock, which evolves in response to CRs
accelerated via DSA as it propagates to the right.
With the normalization constants described above, $\tilde \rho_0=1$, 
$\tilde u_0=-1$, and 
$\tilde P_{\rm g,0}= (1/\gamma_g)M_0^{-2}$ in code units,
where $M_0$ is the accretion flow Mach number. 
The gas adiabatic index, $\gamma_{\rm g}=5/3$.
The shock normalization constants are set by two parameters,
$M_0$ and $c_{s,0}$, through 
\begin{equation}
 u_{\rm n} = |u_0|= c_{s,0} M_0,
\end{equation}
where $c_{s,0} = 15 ~\kms(T_0/10^4)^{1/2}$ is 
the upstream sound speed of the accretion flow. 
We assume the preshock gas is fully ionized with a mean molecular
weight, 0.61. We do not include complexities due to
interactions between neutral and ionized gas in our numerical treatment.
We also ignore added energy transport processes such as radiative cooling, the
diffuse radiation emitted by the shock-heated gas and thermal
conduction. 
However, they should be included in 
future studies for more realistic model calculations. 

For a gasdynamic shock with compression ratio,
$r$, the speed of the shock reflected from 
the piston is $u_s = u_{\rm n}/(r-1)$ in the piston (\ie simulation) frame
and $u_s^\prime=u_{\rm n}r/(r-1)$ in the upstream flow frame.
From this relation, 
the {\it initial} shock speed with respect to the upstream flow
spans the range $u_{s,0}^\prime=(4/3-3/2)u_{\rm n}$
for accretion Mach numbers, $2\le M_0\le 100$. 
However, as CR pressure builds in the shock precursor,
the total compression through the shock increases, and the shock
slows down.
That is, the {\it instantaneous} shock speed, $u_s^\prime(t)$,
relative to the accretion flow decreases over time.
In addition, the gas subshock weakens as the flow is decelerated
and heated in the precursor by the CR pressure gradient.  

We ran most simulations until time $t_{\rm f}/t_{\rm n} =20-40$, 
which is generally sufficient to reach an approximate dynamical
steady state, as discussed in \S 3. Several much longer simulations 
based on a new ``coarse grained momentum finite volume'' code in 
development have also been carried out to confirm the
time asymptotic behaviors, as discussed in \S 3.5.
The simulated space is $x/x_{\rm n}=[0,20]$ for $t_{\rm f}/t_{\rm n}=20$ 
and $x/x_{\rm n}=[0,40]$ for $t_{\rm f}/t_{\rm n}=40$ with 
$N=(1-2)\times10^4$ spatial zones on the base grid.
The simulations were carried out on a base grid with
$\Delta x_0 = (2-4)\times 10^{-3}$ using $l_{\rm max}=7$ additional grid 
levels with a refinement of two,
so $\Delta x_7 = \Delta x_0 / 2^7$ at the finest grid level.
The number of refined zones around the shock is $N_{rf}=50$ on the base
grid and so there are $2N_{rf}=100$ zones on each refined level.
To avoid technical difficulties
the AMR technique is turned on and the CR injection and acceleration are activated
only after the shock propagates to a half length of the pre-defined AMR refinement 
region in the base grid. 
This initial delay of the CR injection and acceleration should not
affect the final outcomes.
For all models we use 230 uniformly spaced logarithmic momentum zones
in the interval $\log (p/m_p c)=[\log p_0,\log p_1]=[-3.0,+3.0]$

Shocks formed by steady accretion flow onto cosmic structure pancakes
provide a convenient prototype of our modeled shocks.
We, therefore, refer to the upstream
flows as ``accretion flows'', and specify the flow speed and
Mach number with respect to the reflecting symmetry plane.
The models are also applicable to other cosmic structure shocks associated
with filaments, knots and individual clusters.
In \citet{kang02} we considered a set of models with fixed infall speed,
$u_0=-1500\kms$ (``constant $u_0$'' models), 
while the preshock temperature was varied
according to $T_0 = 10^4 {\rm K} (100/M_0)^2$ with $2\le M_0\le 100$.
In \citet{kang03}, on the other hand, we 
set the preshock temperature at $T_0=10^4$ K (``constant $T_0$'' models),
and varied the shock speeds according to $u_0= (-15\kms) M_0$,
with $5 \le M_0\le 50$. 
We now extend the study of \citet{kang03} to include higher preshock
temperatures, and also to include the effects of pre-existing
CRs in the shock evolution. 
In particular we include:
1) $T_0=10^6$ K shocks without pre-existing CRs, 
2) $T_0=10^4$ K shocks with pre-existing CRs, and 
3) $T_0=10^6$ K shocks with pre-existing CRs. 
For $T_0=10^4$ K upstream conditions models with $5\le M_0\le 5$ 
( $75\kms \le u_{\rm n}\le 750 \kms$) are considered, 
while for $T_0=10^6$K conditions
we include cases with $2 \le M_0\le 50$ ($300\kms \le u_{\rm n}\le 7500 \kms$).
We note, for $T_0=10^6$K conditions, models with $M_0\le 20$ ($u_{\rm n} \le 
3000\kms$) would be relevant for cosmic structure shocks, but models with
$M_0=30$ and 50 are included to explore the similarity issue 
discussed above in \S2.4. 
In determining these sets of model parameters, we considered the following
factors.
In astrophysical environments photoionized gas of $10^4$ K is
quite common. For example, Ly $\alpha$ clouds and the intergalactic medium
in cosmological sheets and voids are believed to be photoionized during
the reionization epoch \citep{Loeb02}. 
Also the preshock gas can be photoionized and
heated to $\sim 10^4$ K by the diffuse radiation emitted by hot postshock gas
when $u_s \gsim 110 \kms$ 
\citep{Shull79, kangshapiro92, Draine93}.
On the other hand, hot and ionized gas of $>10^6$ K is also found in 
the hot phase of the ISM 
\citep{McKee77, Spitzer90} and in the intracluster medium of X-ray
clusters \citep{Fabian94}. 

Since cosmological simulations indicate that most diffuse plasma in 
large scale cosmic structures is
likely to be shocked more than once during structure formation, it is 
likely that many ``internal shocks'' process plasma
that carries previously accelerated CRs. We, therefore, include
shock simulations including these populations.
For pre-existing, upstream CRs
we assumed a simple power-law spectrum smoothly connected to the upstream
thermal Maxwellian distribution ($f_M$) at a momentum $p_M$ as 
\begin{equation}
f_{up}(p) = f_M(p_M) (p/p_M)^{- q_{\rm in}}
\end{equation}
for $p_M < p < p_1$,
where $p_M= 5(m_p k_B T_0)^{1/2}$, $p_1=10^3$. 
For $T_0=10^4$ K shocks we considered in the pre-existing
CR cases an upstream CR pressure
condition, $P_{\rm c,0} =0.25 P_{\rm g,0}$, using $q_{\rm in}=4.7$.
For $T_0=10^6$ K shocks we included three different upstream
conditions; namely, $P_{\rm c,0} =0.25 P_{\rm g,0}$ ($q_{\rm in}=4.5$),
$P_{\rm c,0} = 0.5 P_{\rm g,0}$ ($q_{\rm in}=4.4$) and
$P_{\rm c,0} = P_{\rm g,0}$ ($q_{\rm in}=4.3$).
The CR spectral forms represent test-particle populations for
previous shocks with moderate Mach numbers in the range, $2.5 \lsim M_s \lsim 4$.
Table 1 provides a summary of the
new models computed for the present study.

\subsection{Injection and Acceleration Efficiencies} 
It is useful to characterize the efficiency with which low energy CRs are
injected at the shocks.
We define the injection efficiency as the fraction of particles that have
entered the shock from far upstream and then injected into the CR distribution:
\begin{equation}
\xi(t)=\frac {\int_0^{x_{max}} {\rm d x} \int_{p_{\rm inj}}^{p_1} 4\pi f_{\rm CR}(p,x,t)
p^2 {\rm d p}}
{ \int_{t_i}^t n_0 u_s^\prime (t^{\prime}) {\rm d t^{\prime}} }\,
\label{xieq}
\end{equation}
where $f_{\rm CR}$ is the CR distribution function,
$n_0$ is the particle number density far upstream,
and $t_i$ is the time when the CR injection/acceleration is turned on
($t_i \approx 3 t_n$ for $M_0\ge 5$ and $t_i \approx 2 t_n$ for $M_0=2$).

If the subshock speed becomes steady, then $n_0 u_s^\prime$ is the same as the
particle flux swept by the subshock, $n_1 u_{\rm sub}$, where
$n_1$ is the particle number density immediately upstream of the subshock.
However, in our simulations these two flux measures can differ by up 
to 10\% for strong
shock models, because the shock speed is changing slowly.

As a measure of efficiency of CR energy extraction at shocks,
we define the ``CR energy ratio'', $\Phi$; namely, the
ratio of the total CR energy within the simulation box to
the kinetic energy in the {\it initial shock frame}
that has entered the simulation box from far upstream,
\begin{equation}
\Phi(t)=\frac {\int_0^{x_{max}} {\rm d x} E_{\rm CR}(x,t)}
 {  0.5\rho_0 (u_{s,0}^{\prime})^3  t }.
\label{crenrat}
\end{equation}
All shock models have the same normalized upstream mass density and velocity,
$\rho_0$ and $u_0$, but different upstream gas thermal energy density, 
$P_{\rm g,0}/(\gamma_g - 1)$, depending on $M_0$.
Therefore, we use the kinetic energy flux rather than the total
energy flux to normalize the ``CR energy ratio''.
Alternatively, the ratio defined with the total energy flux,
$ F_{\rm tot}=u_{s,0}^{\prime}[0.5\rho_0 (u_{s,0}^{\prime})^2 + 
\gamma_g P_{g,0}/(\gamma_g-1) +  \gamma_{c,0} P_{c,0}/(\gamma_{c,0}-1)]$,
 can be calculated from the following relation, 
\begin{equation}
\Phi_{tot}(t) \approx { \Phi(t) \over {1+ {3 \over M_0^2} 
({u_0\over u_{s,0}^{\prime} })^2
(1+ P_{c,0}/P_{g,0})}},
\label{crenprime}
\end{equation}
where we assume $\gamma_{c,0}\approx 5/3$ for the preshock CR population.
This shows that $\Phi_{tot} \approx \Phi$ for large values of $M_0$,
while $\Phi_{tot} \approx (3/5) \Phi$ for $M_0=2$ and $P_{c,0}=P_{g,0}$.
\section{Simulation Results}

We can define three key stages of shock evolution in response to CR feedback:
1) Development of the shock precursor that slows and heats the
flow entering the gas subshock, reducing the Mach number of the latter;
2) achievement of an {\it approximately} time-asymptotic dynamical
shock transition, including nearly steady postshock CR and gas 
pressures; 
3) continued, approximately ``self-similar'' 
evolution of the shock structure for Bohm-type diffusion, 
as CRs are accelerated to ever higher momenta.
Once stage (3) is reached, there is little change in the dynamical
properties of the shocks, and, in particular, little change in the
total efficiency with which kinetic energy is transferred to CRs.
Thus, given the rapidly increasing computational cost
to simulate acceleration of CRs to still higher energies
with Bohm diffusion (which more than quadruples for each
factor of two increase in maximum CR momentum), we deemed it 
sufficient to the questions being
addressed to terminate the simulations at the times indicated. 

As \citet{kang03} showed,  the evolution of 
CR modified shocks with Bohm-type diffusion are
mainly determined by the shock Mach number for a given injection parameter. 
This includes the efficiency of energy transfer to CRs, $\Phi$.
That is, two shocks with the same Mach number, but with different shock
speeds evolve qualitatively similarly 
when expressed in terms of characteristic diffusion scales, $t_n$ and
$x_n$.
During the early evolutionary stage, however, the similarity 
between the two shocks may break down as pointed out in \S 2.4. 
As described there, this happens
if the shock speed is larger than a percent or so of the speed of light, 
on account of the different momentum-velocity scalings 
for nonrelativistic and relativistic particles at injection.
As a result, the ratio of $P_c/(\rho_{\rm n} u_{\rm n}^2)$ increases faster 
and nonlinear modification sets in earlier in terms of $t/t_{\rm n}$ 
in the shock with lower speed.
Thus, depending on when the transition from nonrelativistic to relativistic
population occurs, early evolution of the two shocks can be different
(\eg see $M_0=30$ models in figure 2), although they still
eventually approach similar time-asymptotic states. 

The other factor that can affect the similarity is the presence of
pre-existing CRs.  For strong shocks of $M_s > 10$, the preshock CR pressure
is only a few percent of the total energy flux even when
$P_c \sim P_g$ far upstream.
In those cases the presence of pre-existing CRs is effectively 
equivalent to having a slightly 
higher injection rate (\ie larger $\epsilon$), which speeds up
the initial shock evolution. 
For strong shocks, our previous studies have shown 
that the time asymptotic CR acceleration efficiency depends only weakly on
the injection rate \citep{kangetal02, kang02},
provided the injection remains above
a minimum threshold for ``efficient'' DSA \citep[see also, \eg][]{bek96}. 
On the other hand, for weak shocks, a pre-existing CR pressure comparable 
to the upstream gas pressure represents a
significant fraction of the total energy entering the
shock, so pre-existing CRs have far more impact.  
Also the time asymptotic CR acceleration efficiency in weak shocks depends 
sensitively on the injection rate and increases with $\epsilon$.
Hence we should expect that relatively weak CR shocks ($M_s<5$) will
be substantially altered by the presence of a finite upstream $P_c$. 

\subsection{Evolution of Shock Structures}

Figure 1 compares the time evolution of two CR modified shock models
with accretion Mach number $M_0=5$, but with different physical
flow speeds and upstream CR conditions. 
The light lines show the flow model with $T_0=10^4$ K 
and the pre-existing CR population, $f_{up} \propto p^{-4.5}$,
while the heavy lines show the model with 
$T_0=10^6$ K and $f_{up} \propto p^{-4.7}$.  
The pressure due to the pre-existing CR particles is 
$P_{c,0}=0.25P_{g,0}$ in both models.
Note that the physical time and length scales are different, since
$t_{\rm n}\propto u_{\rm n}^{-2}$ 
and $x_{\rm n}\propto u_{\rm n}^{-1}$ 
($u_{\rm n} = 75 ~\kms$ for $T_0=10^4$ K models
and $u_{\rm n} = 750 ~\kms$ for $T_0=10^6$ K models).

In the model with $u_{\rm n}=75\kms$ 
CR particles are injected at much lower injection momenta.
Since for a fixed Mach number their initial acceleration times scale 
as $\tau_{acc}/t_{\rm n} \propto p_{\rm inj}^2$, the acceleration
and shock modification takes place more rapidly in these
time units compared to the model with $u_{\rm n}=750\kms$.
The value of $P_c/(\rho_{\rm n} u_{\rm n}^2)$ approaches 0.8
for the model with $u_{\rm n}=75\kms$, while it approaches 0.65
for the model with $u_{\rm n}=750\kms$. 
Otherwise, the overall evolution is roughly similar. 

Figure 2 shows that even for much higher Mach numbers, $M_0=30$, 
two shocks with $u_{\rm n}=450 ~\kms$ and $u_{\rm n}=4500 ~\kms$
approach similar time-asymptotic states, 
although their early evolutions are very different.
However, as the shock speed increases
further, substantial differences develop in the early evolution.
For $M_0=50$ models, for example, 
unlike the shocks shown in the figures,
the overall shock structures then evolve much more slowly in
our normalized time units, since the injection momentum is no
longer nonrelativistic.
Then the Bohm-like diffusion coefficient and the
CR partial pressure near this
momentum do not scale with $p^2$ as they do for nonrelativistic momenta.
The approximate self-similarity of the shock evolution breaks
down under those conditions.
Even those shocks, however, do approach similar time
asymptotic states for a given Mach number.

The flow entering the piston is decelerated first by the precursor and
then by the subshock in our simulation.
For a pure gas fluid with a constant adiabatic index, the postshock flow
speed relative to the piston should be uniformly zero.
This should be approximately true for the models with inefficient
CR acceleration in our simulations (see, for example, Fig.4 below).
In CR modified shocks, however, the ``effective'' adiabatic index
decreases due to increasing CR energy, so the total compression
through the shock increases and the shock slows down over time.
This transition in the compressibility of the flow leads to a
net forward flow velocity of $0.04<u/u_{\rm n}<0.07$ in the piston rest
frame for the two models shown in Figs. 1 and 2.
However, the net flow speed is decelerated by the positive gradient
of the combined postshock pressure ($P_c+P_g$) and approaches a smaller
value ($0.02<u/u_{\rm n}<0.03$) at the terminal simulation time.

As noted in previous {\it time-dependent} studies (see the references
in \S 1), the compression factor for these CR dominated shocks
relaxes to values still higher than that for a gasdynamic shock,
even a for a fully relativistic gas, even though energy is
not leaving the system in our simulations. The reason has to do
with the {\it evolving} distribution of energy in the system.
In particular, for a diffusion coefficient increasing with $p$,
especially for high Mach number, CR dominated shocks,
the compression factor can remain  much higher
than what is expected in a steady-state, energy-conserving shock
jump for relativistic gas (\ie $\rho_2/rho_0=7$).
This is because energy is being continuously transported out
of the region of the subshock by diffusion of the highest energy
CRs with an ever-increasing diffusion scale $x_d(p_{\rm max}$).
More extensive discussion can be found in \citet{kangetal02}.

\subsection{Effect of Pre-existing CRs for Weak Shocks}

Figure 3 compares the evolution of shocks {\it with} and {\it without} 
pre-existing, upstream CRs. Four sets of shocks are illustrated:
1) $M_0=5$, $u_n=75~\kms$, and $T_0=10^4$ K (upper left panel),
2) $M_0=5$, $u_n=750~\kms$, and $T_0=10^6$ K (upper right panel),
3) $M_0=30$, $u_n=450~\kms$, and $T_0=10^4$ K (lower left panel), and
4) $M_0=30$, $u_n=4500~\kms$, and $T_0=10^6$ K (lower right panel).
The approximate self-similarity for the evolution of these
shocks, when described in normalized units is apparent.
In addition one can see that the 
presence of pre-existing CRs is clearly more important 
for low Mach number shocks.

It is especially important to consider the influence of 
pre-existing CRs in low Mach number shocks, 
since internal shocks with $M_s\sim 3$ may be the most important for
dissipation in shock-heated gas inside nonlinear structures \citep{ryu03}. 
We already know that
CR acceleration efficiency in low to moderate strength shocks 
depends sensitively on the injection rate and increases with $\epsilon$
\citep{kang02}.
Our new simulations show that the
CR acceleration efficiency in these shocks
depends similarly on the pre-existing (upstream) CR pressure.
That trend was apparent in figure 1.
Figure 4 emphasizes the point impressively
for even lower Mach number shocks. It shows the time evolution of the 
models with accretion flow Mach number $M_0=2$ and $u_0=-300 \kms$, 
which drive shocks with initial Mach number of $M_s=3$ into 
the preshock gas with $T_0=10^6$ K. 
Three models are compared in the figure: 1) a model with 
no pre-existing CRs (solid lines),
2) a model with $P_{c,0}=0.25 P_{g,0}$ (dotted lines), 
and 3) a model with $P_{c,0}=0.5 P_{g,0}$ (dashed line).
For the model without the pre-existing CRs and with injection
parameter, $\epsilon=0.2$, the CR injection is small and the
resulting CR acceleration
is inefficient. The solution is essentially a test-particle
solution, with minimal CR modifications to the flow structure.
There are a couple of related reasons for the impact of the upstream CRs 
in the two other models.
First, the pressure of the pre-existing CRs
cannot be neglected in the shock transition compared to the total 
shock energy flux.
In addition, the inflowing CRs act like a large source of injected
CRs, effectively enhancing the injection rate \citep[see also][]{blasi}.
\citet{kang02} also considered similar $M_0=2$ models without the 
pre-existing CRs ($u_0=-1500\kms$, and $T_0=2.5\times 10^7$ K), 
but with a range of injection parameters, $\epsilon=0.2-0.4$.
We found that the postshock CR pressure increased over this
range of $\epsilon$ from
$P_c = 0.04$ to $P_c = 0.24$, while the associated injection
rate rose from $\xi=10^{-4}$ to $\xi = 10^{-2}$ 
\citep[see figure 8 in][]{kang02}.
These results imply that even in low Mach number shocks
the CR acceleration can have significant dynamical influences, 
when the injection rate is as high as $\xi\sim 10^{-2}$ or
when the upstream CR pressure is greater than a few times 10 \% of the gas 
pressure. 
This could be important
in the context of cosmic structure shocks, especially for 
``internal shocks'' in which the upstream plasma already  
carried a significant population of CRs produced in an earlier era.

\subsection{The Particle Distribution} 
Time evolution of the distribution function $g(p)=f(p)p^4$ {\it at the 
gas subshock} is shown in  figure 5 for the models with $T_0=10^6$ K
and with $f_{up} \propto p^{-4.5}$.
The dotted line shows the initial Maxwellian distribution with $T_0$ and
the specified power-law distribution for the pre-existing CR population.
The postshock thermal distribution during the early evolution
(before shock modification) should have $T_2 \propto M_0^2$,
so the injection momentum, $p_{\rm inj} \propto M_0$ for $t/t_{\rm n} < 1$.
However, as energy transfer to CRs increases over time and the
gas subshock weakens,
the peak in the Maxwellian distribution shifts to lower momenta.
The model with $M_0=30$ shows this effect clearly. 
This nonlinear feedback is greater for higher Mach number, 
so that the postshock temperature of evolved CR modified shocks depends on
$M_0$ much more weakly than the standard $M_0^2$. 

It is useful to identify $p_{\rm max}$ as the momentum above which $g(p)$ drops
sharply, characterizing the effective upper cutoff in the CR
distribution, at a given time since shock formation. 
This momentum is generally easily identified.
We find that it is approximately related to the age of the shock
as $p_{\rm max} \sim 0.4 (t/t_{\rm n})$ for $p_{\rm max}>1$.
This explains why values of $p_{\rm max}$ are
similar at a given value of $t/t_{\rm n}$ for all models considered here,
independent of $u_{\rm n}$, injection momentum, or shock Mach number.
The previously described back reaction of the CRs on the gas subshock
acts to move the shock towards a limiting form and postshock
state.
On the one hand, a fraction of CR particles injected at early times
continues being
accelerated to higher momenta, so that $p_{\rm max}(t)$ always
increases until they are subject to escape due to
geometry or insufficient scattering. On the other hand, 
particle injection becomes less efficient as the subshock feature weakens.
Because of this reduced source spatial
diffusion causes the number density of CRs near the subshock
at a given momentum 
to decrease with time for any $p<p_{\rm max}(t)$.
Combined, these effects cause the postshock CR pressure, $P_c$, to
approach a time-asymptotic value,
and, for Bohm-like diffusion, the shock structure to become approximately self-similar.

\subsection{Cosmic Ray Injection and Acceleration Efficiencies}

Figure 6 illustrates the time averaged injection efficiency, $\xi(t)$, defined
by equation (9) for models without pre-existing CRs.
The calculated injection efficiency ranges from $\sim~{\rm a~few}~\times 10^{-5}$ to 
$\sim 10^{-3}$, depending on the subshock Mach number,
the shock speed, and the injection parameter, $\epsilon$.
As outlined in \S 2.4 and discussed fully in \citet{kangetal02}, the 
modeled thermal-leakage injection
process is less efficient for weaker subshocks
and for smaller $\epsilon$ (stronger wave trapping of suprathermal
particles in the postshock environment).
Injection is also less efficient for lower shock speeds,
because the diffusive flux of injected particles crossing the subshock
is smaller for smaller injection momentum ($p_{\rm inj} \propto u_s$).
So, for a given Mach number the models with higher shock speed (\ie higher
$T_0$ in figure 6) have larger values of $\xi(t)$.
For strong shocks the subshock weakens significantly due to the
CR nonlinear feedback, so the injection efficiency slowly decreases over time. 

Figure 7 shows for the models with $T_0=10^4$K the CR energy ratio, $\Phi$, 
the CR pressure and the gas pressure at the subshock
normalized by the far upstream ram pressure in the {\it initial} shock frame, 
\ie $P_{\rm c,2}/(\rho_0 u_{s,0}^{\prime~ 2})$,
and $P_{\rm g,2}/(\rho_0 u_{s,0}^{\prime~ 2})$.
In addition the bottom panels show the CR pressure normalized 
by the far upstream ram pressure in the {\it instantaneous} shock frame,
$P_{\rm c,2}/(\rho_0 u_{s}^{\prime~2})$,
The left panels illustrate the models without pre-existing CRs,
while right panels show the models with $P_{c,0}=0.25 P_{g,0}$. 
As described in the previous subsection, the postshock $P_c$  for
all Mach numbers increases until a balance between injection/acceleration and 
advection/diffusion of CRs is achieved. After that time, 
which corresponds in these simulations to the generation of moderately 
relativistic CRs in the shock, the postshock CR pressure remains almost steady.
Simultaneously, the CR energy ratio, $\Phi$, also
initially increases with time, but then asymptotes to
a constant value, once $P_{\rm c,2}$ has reached a  quasi-steady value.
The asymptotic values result from the ``self-similar'' behavior of the 
$P_{\rm c}$ distribution. 
Since both the numerator and denominator in equation (\ref{crenrat}) increase 
approximately linearly with time in the case of the self-similar evolution 
of $E_{\rm c}$, $\Phi$ becomes steady when $P_{\rm c,2}$ becomes constant.
As illustrated in Figure 7, time-asymptotic values of $\Phi$ increase 
with the accretion Mach number $M_0$, approaching
$\Phi\approx 0.5$ for $ M_0>20$ by the simulation termination times.
As expected from the discussion in \S 3.4, the presence of 
pre-existing CRs increases the asymptotic postshock
$P_c$ for the weaker shock, $M_0=5$ model, while it does not affect the 
strong shock models with $M_0\ge 10$.
Figure 7 also illustrates the well-known result that strong shocks 
can be mediated mostly by CRs
while the postshock gas thermal energy is reduced greatly compared to that in
conventional gasdynamic shocks of the same Mach number. 
As a result of nonlinear feedback the shock slows down and so the
ratio of $P_{\rm c,2}/(\rho_0 u_s^{\prime~2})$ can reach $\sim 0.9$,
while the ratio of $P_{\rm c,2}/(\rho_0 u_{s,0}^{\prime~2})$ reaches 
$\sim 0.6$ in these models.

In figure 8 we present the same quantities for the models with $T_0=10^6$K.
As explained in the previous section, due to the higher injection momenta, 
the CR acceleration is slower to develop for the models 
with $M_0\ge 20$, compared to the same Mach number model with lower shock speed.
Especially for the $M_0=50$ model, the postshock $P_c$ continues to increase at the
termination time ($t/t_o=40$) of the simulation.
Again, the presence of pre-existing CRs significantly increases the postshock $P_c$
at low Mach number models ($M_0\le 5$).
We show the same quantities for low Mach number simulations with even higher pre-existing 
$P_c$ in figure 9.
We see that the postshock $P_c$ increases with the pre-existing $P_c$ for 
these low Mach number models. 

\subsection{Time Asymptotic Behaviors}
As discussed above, some characteristics of the CR modified shocks seem to 
have reached quasi-steady states well before the termination time 
for most of the models considered in this work.
We have further explored these time asymptotic behaviors by performing much 
longer simulations with a new 
``Coarse-Grained Momentum Finite Volume'' (CGM) numerical scheme 
under development. CGM utilizes the
fact that the distribution function can be approximated as a piecewise
power-law within a given broad momentum bin and so only roughly one
momentum bin per octave is
necessary to follow the CR population numerically. Details and tests
of the 
CGM numerical method will be presented in a forthcoming paper \citep{twj04}. 
In figures 10 and 11, we show the simulation results of $M_0=5$ and $M_0=30$
models, respectively, which were calculated by this CGM method up 
to $t/t_{\rm n}=1000$. 
No pre-existing CRs were assumed and 14 momentum bins were used for these
calculations.
The heavy solid lines are for the simulation results 
at $t/t_{\rm n}=20$ for the $M_0=5$ model 
and $t/t_{\rm n}=40$ for the $M_0=30$ model 
which were calculated by the ``fine-grained momentum'' method described 
in \S 2 and used for all the other results presented in this paper. 
Those simulations were carried out with 240 momentum bins. 
These figures confirm very firmly the aforementioned "self-similarity" 
of the shock evolution and the time-asymptotic behaviors. 

Thus we show in figure 12 values of the CR energy ratio $\Phi$ 
at the termination time 
as a function of the initial unmodified shock Mach number, $M_s$, 
as a simple measure of the CR acceleration efficiency.
For convenience in comparing with other plots
Values of $M_s$ are listed against $M_0$ in Table 1. 
Flows with $T_0=10^4$ are shown in the upper panel, while
shocks developed in $T_0=10^6$ flows are represented in the
lower panel.
For comparison, we also plotted in the upper panel 
results from ``constant $u_0$'' models with
$u_0= -1500 \kms$ and $T_0=10^4 (100/M_0)^2$ K, originally presented in
figure 7 of \citet{kang03}. 
Those models, which did not include
a pre-existing CR population, behave similarly to the new
models considered in the present work.
In each series of simulations the efficiency asymptotes to $\Phi \approx 0.5$
for large $M_s$. This asymptotic efficiency is independent of
$P_{c,0}/P_{g,0}$; \ie
the level of pre-existing CRs.
It also reflects the fact that the postshock CR pressures in the high 
Mach number shocks evolve in all cases roughly to 60\% of the
inflowing ram pressure measured in the initial shock rest 
frame, $\rho_0u_{s,0}^{\prime ~2}$.
In weak to moderate strength shocks, on the other hand, the CR 
efficiencies shown in figure 12, or the postshock CR pressures 
illustrated in figures 7 and 8, depend strongly on $P_{c,0}/P_{g,0}$.
A rough dividing line
for these two behavior domains is $M_s \sim 20$. 
For the $M_s=3$ model without the pre-existing CRs and $\epsilon=0.2$, 
the injection rate $\xi \approx 2 \times 10^{-5}$ and
$\Phi$ approaches to only 0.04.
Since $0.3\lsim \epsilon \lsim 0.4$ is indicated for strong shocks
\citep{malvol98} and weaker wave fields are expected in lower Mach shocks, 
leading to larger values of $\epsilon$, 
we show three addition models for $\epsilon=$ 0.25, 0.3, and
0.4 with $M_s=3$, $T_0=2.5\times 10^7$K and no pre-existing CRs,
originally presented in figure 8 of \citet{kang02}, 
in the upper panel of figure 12.
In \citet{ryu03} the CR efficiency model, $\Phi(M_s)$, for $\epsilon=0.3$
was adopted in the calculation of the CR energy generated by cosmic
stricture shocks.
We note that the functional minimum in $\Phi(M_s)$ for the models with
pre-existing CRs is an artifact of the definition of $\Phi$,
which is normalized by the inflowing kinetic energy, not the total
energy. The thermal energy flux is a significant fraction of the
kinetic energy flux for low Mach number shocks.
There is no minimum in the function $\Phi_{tot}$, as defined
in equation \ref{crenprime}. 
For instance the value
$\Phi(M_s = 3) \approx 0.67$ shown in figure 12, becomes $\Phi_{tot}(M_s = 3)
\approx 0.4$.

As also pointed out in \S 3.2, the increases in CR energy ratio 
for weak shocks when they process plasma that
contains pre-existing CRs, could be important to understanding the
properties of cosmic structure formation shocks. 
In addition, most baryonic matter in clusters
has been shocked more than once before the current epoch \citep{ryu03}.
This means that many of the shocks primarily responsible for
energy dissipation during structure formation may be influenced
by CRs accelerated in previous shock events, since magnetic field
strengths at levels discussed in the introduction should
tie CRs up to moderately relativistic energies to the thermal
plasma over cosmic time scales \citep[\eg][]{volk96}. 
In that case, those shocks may be substantially more efficient at 
CR acceleration than one might otherwise expect.

\section{Summary}

Energy dissipation at collisionless shocks involves complex physical processes
and plays crucial roles in the evolution of shock-heated astrophysical plasmas. 
Here our focus is CR acceleration via the first-order Fermi process
at cosmic shocks that form in and around nonlinear structures 
during the formation of large scale structure in the universe. 
Using our cosmic-ray AMR Shock code \citep{kangetal02},
we have calculated the CR acceleration at 1D quasi-parallel shocks
driven by plane-parallel infall flows with $|u_0|=75-7500 {\rm km s^{-1}}$ 
and preshock temperatures of $T_0=10^4-10^{7.6}$K.
Mach numbers of the resulting shocks range over $ 2\le M_s\le 133$,
which covers the characteristic shock Mach numbers found in cosmological 
hydrodynamic simulations of a $\Lambda$CDM universe \citep{ryu03}.

The main purpose of this study is to explore how the CR acceleration
efficiency at cosmic structure shocks depends on the preshock temperature, 
shock speed and pre-existing CRs as well as shock Mach number.
This, in turn, affects the thermal and dynamical history of the shocked-heated baryon 
gas, since the energy transfer to CRs reduces the thermal energy of the gas
and increases the compressibility of the gas flow, much like
radiative shocks.

Unlike pure gasdynamic shocks, 
downstream states of the CR modified shocks cannot be found from
"shock jump conditions", so they have to be integrated
time-dependently from given initial states or found from nonlinear
analytical methods.
Time-dependent simulations are computationally expensive,
because the CR diffusion and acceleration processes contain 
a wide range of length and time scales. 
Fortunately, there are approximate similarity properties that can be 
employed to study the time asymptotic behavior of evolved CR modified shocks
when they are mediated by Bohm-like diffusion.  
Firstly, the CR pressure approaches a steady-state value
in a time scale comparable to the acceleration time scales
for mildly relativistic protons after which
the evolution of CR modified shocks becomes approximately ``self-similar''.
This feature enables us to predict time asymptotic values of the CR
acceleration efficiency, although we followed the CR 
acceleration up to only moderately relativistic energies
(\ie $p_{\rm max}/(mc)\sim 16$).
We have confirmed these time asymptotic behaviors through much longer 
simulations with a new, more efficient, numerical scheme \citep{twj04},
which were extended to achieve the cut-off momentum $p_{\rm max}/(mc)\sim 400$. 
Secondly, 
{\it two models with the same Mach number, but with different
accretion speeds} show qualitatively similar evolution and
structure
when the dynamical evolution is presented in terms of characteristic
diffusion scales, $t_{\rm n}=\kappa_{\rm n}/u_{\rm n}^2$ and 
$x_{\rm n}=\kappa_{\rm n}/u_{\rm n}$.
Such similarities are only approximate, however, 
because the partial CR pressure and the Bohm diffusion coefficient
for transrelativistic CRs do depend on the momentum $m_pc$.
Since the effective injection momentum is $p_{\rm inj}/m_p c \propto (u_s/c)$,
the initial evolution depends on the shock speed as well as Mach number.

Finally, we state the main conclusions of our study, which can be summarized as follows:

1) Suprathermal particles can be injected efficiently
into the CR population at quasi-parallel cosmic shocks via the thermal leakage process.
For a given injection parameter defined in the text, $\epsilon$, 
the fraction of injected CRs 
increases with the subshock Mach number, but approaches $\xi \sim 10^{-3}$
in the strong shock limit. 

2) For a given value of $\epsilon$, the acceleration efficiency 
increases with the shock Mach number, but approaches a similar value
in the strong shock limit. 
Time asymptotic values of the ratio of CR energy to inflowing kinetic
energy, $\Phi$, converge to $\Phi\approx 0.5$ for $M_s \gsim 30$
and it is relatively independent of other upstream or injection parameters
(see figure 12). 
Thus, strong cosmic shocks can be mediated mostly by CRs and the gas 
thermal energy can be up to $\sim 10$ times smaller than that expected 
for gasdynamic shocks. 

3) For weak shocks, on the other hand, the acceleration efficiency increases
with the injection rate (or $\epsilon$) and the pre-exiting $P_c$ (see figure 12). 
For cosmologically common $M_s=3$ shocks, for example, $\Phi \sim 0.04$ for $\epsilon=0.2$ and 
$\Phi \sim 0.2$ for $\epsilon=0.3$ in the absence of pre-existing CRs.
The presence of a pre-existing CR population acts effectively as a
higher injection rate than the thermal leakage alone, leading
to greatly enhanced CR acceleration efficiency in low Mach number shocks. 
We found even for weak shocks of $M_s=3$, that up to 40 \% of the total
energy flux through the shocks can be transferred to CRs, when
the upstream CR pressure is comparable to the gas pressure in the preshock flow. 

\acknowledgements
HK was supported by KOSEF through Astrophysical Research Center
for the Structure and Evolution of Cosmos (ARCSEC) and grant-R01-1999-00023
and by Pusan National University Research Grant.
TWJ is supported by NSF grants AST00-71167 and AST03-07600, 
by NASA grant NAG5-10774
and by the University of Minnesota Supercomputing Institute.

\clearpage

\begin{figure}
\plotone{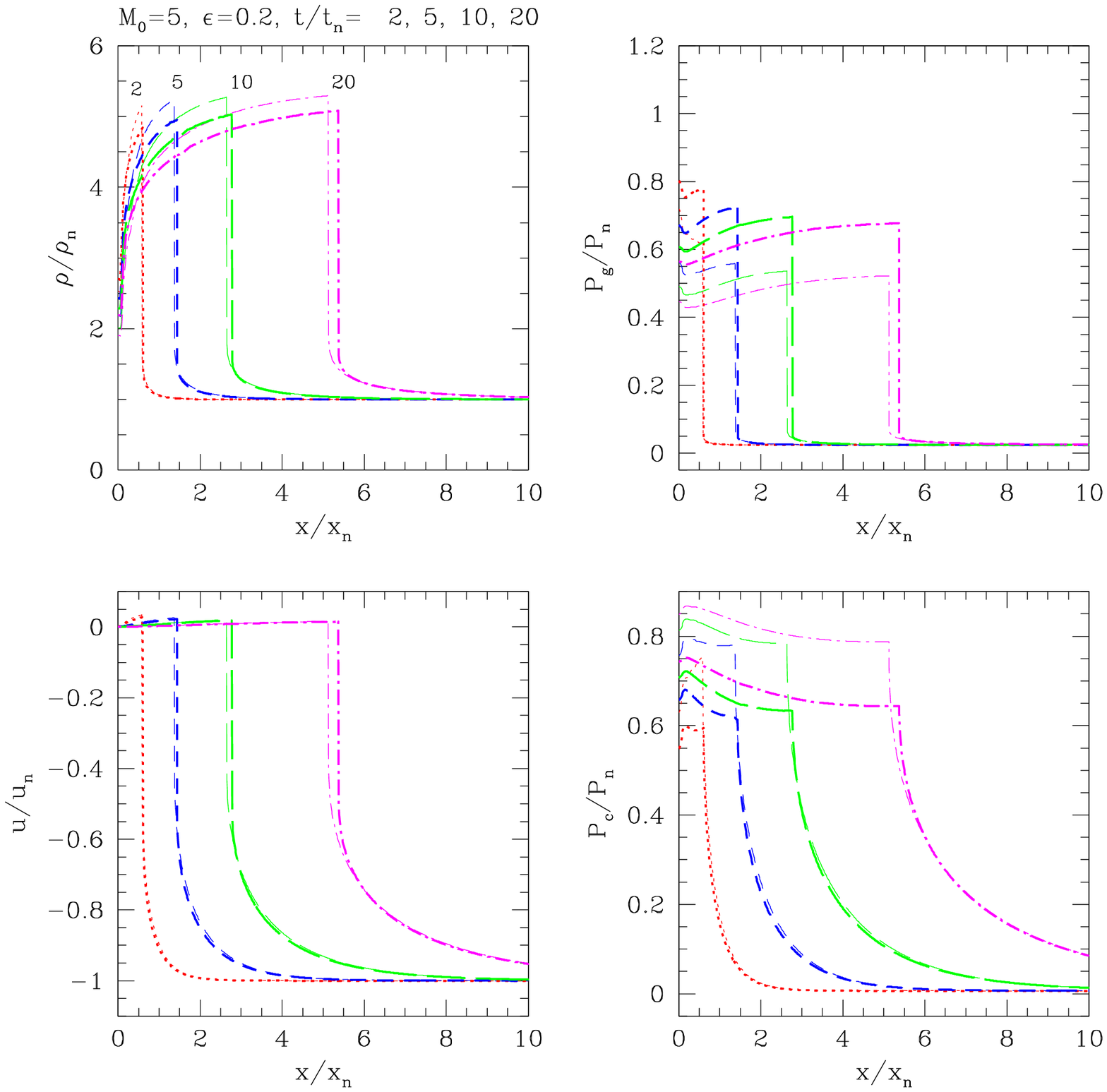}
\figcaption{
Time evolution of the shocks driven by 1D accretion flows with
$M_0=u_{\rm n}/c_{s,0}=5$ is shown at normalized times, $t/t_{\rm n}=$ 2, 5, 10, and 20.
The accretion flow enters from the right boundary ($x/x_{\rm n}=20$)
and is reflected at the piston ($x/x_{\rm n}=0$),
leading to a shock with $M_s=6.8$ propagating to the right.
The flow speed is shown in the piston rest frame, so the flow is almost
at rest near the piston. 
The leftmost profile corresponds to the earliest time.
Light lines represent a flow with $u_{\rm n}=75~\kms$, $T_0=10^4$ K,
and a pre-existing CR population with $f_{up} \propto (p/p_M)^{-4.7}$.
Heavy lines represent the model with $u_{\rm n}=750\kms$, $T_0=10^6$ K,
and $f_{up} \propto (p/p_M)^{-4.5}$.
For both models the upstream CR pressure is $P_{c,0}/P_{g,0}=0.25$.
The normalization diffusion time scale, $t_{\rm n}=\kappa_{\rm n}/u_{\rm n}^2$,
and diffusion length, $x_{\rm n}=\kappa_{\rm n}/u_{\rm n}$ are
defined by the accretion speed of each model.
The inverse wave-amplitude parameter $\epsilon=0.2$ is adopted for
both models.
\label{fig1}}
\end{figure}
\clearpage

\begin{figure}
\plotone{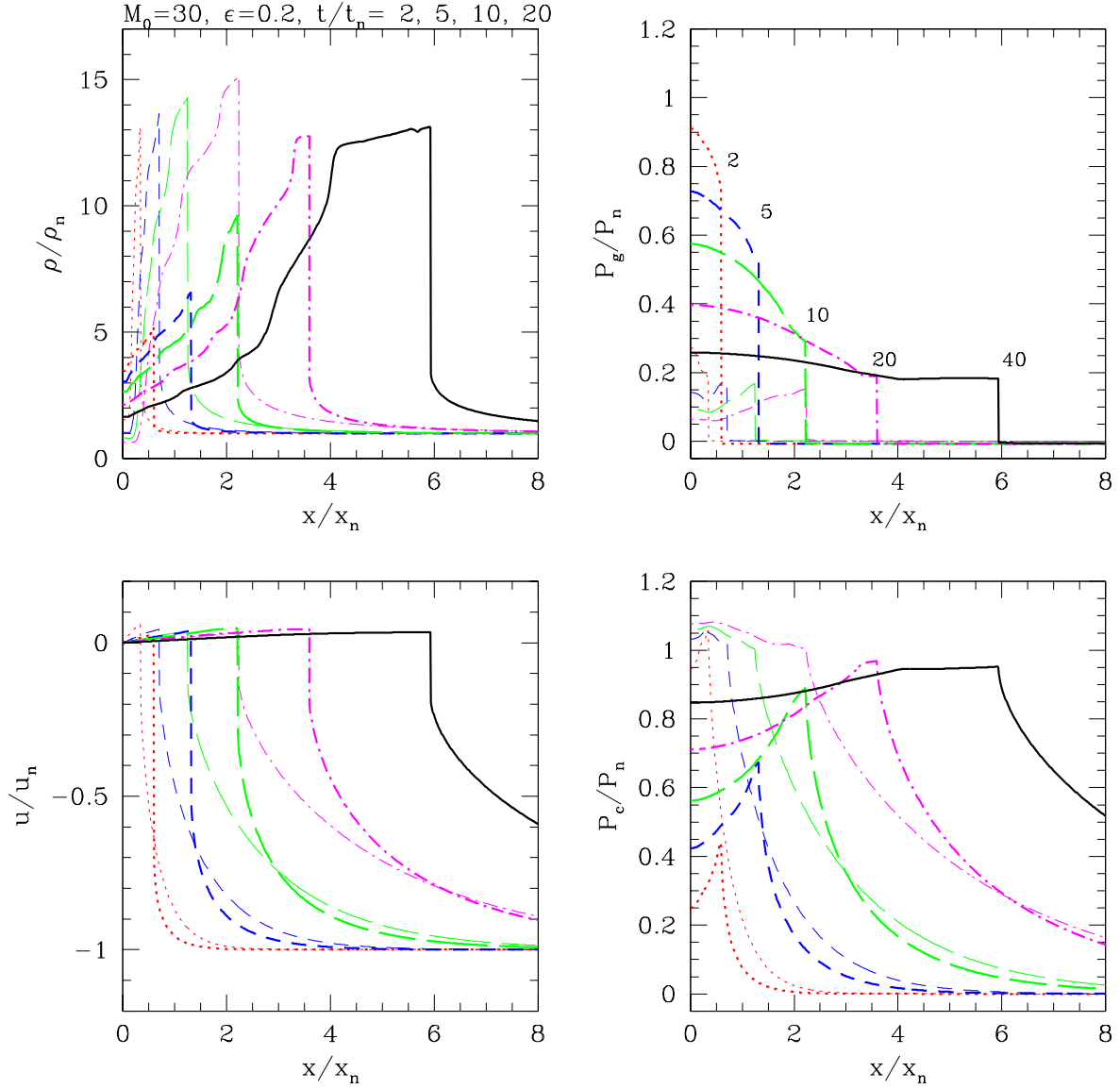}
\figcaption{
Same as Figure 1 except $M_0=30$.
Light lines are for the model with $u_{\rm n}=450~\kms$, $T_0=10^4$ K,
and a pre-existing CR population of $f_{up} \propto (p/p_M)^{-4.7}$, while 
heavy lines represent the model with $u_{\rm n}=4500\kms$, $T_0=10^6$ K,
and $f_{up} \propto (p/p_M)^{-4.5}$.
For both models the upstream CR pressure is $P_{c,0}/P_{g,0}=0.25$.
Note for $T_0=10^6$K model additional curves (heavy solid lines) are
shown at $t/t_{\rm n}=40$.
\label{fig2}}
\end{figure}
\clearpage

\begin{figure}
\plotone{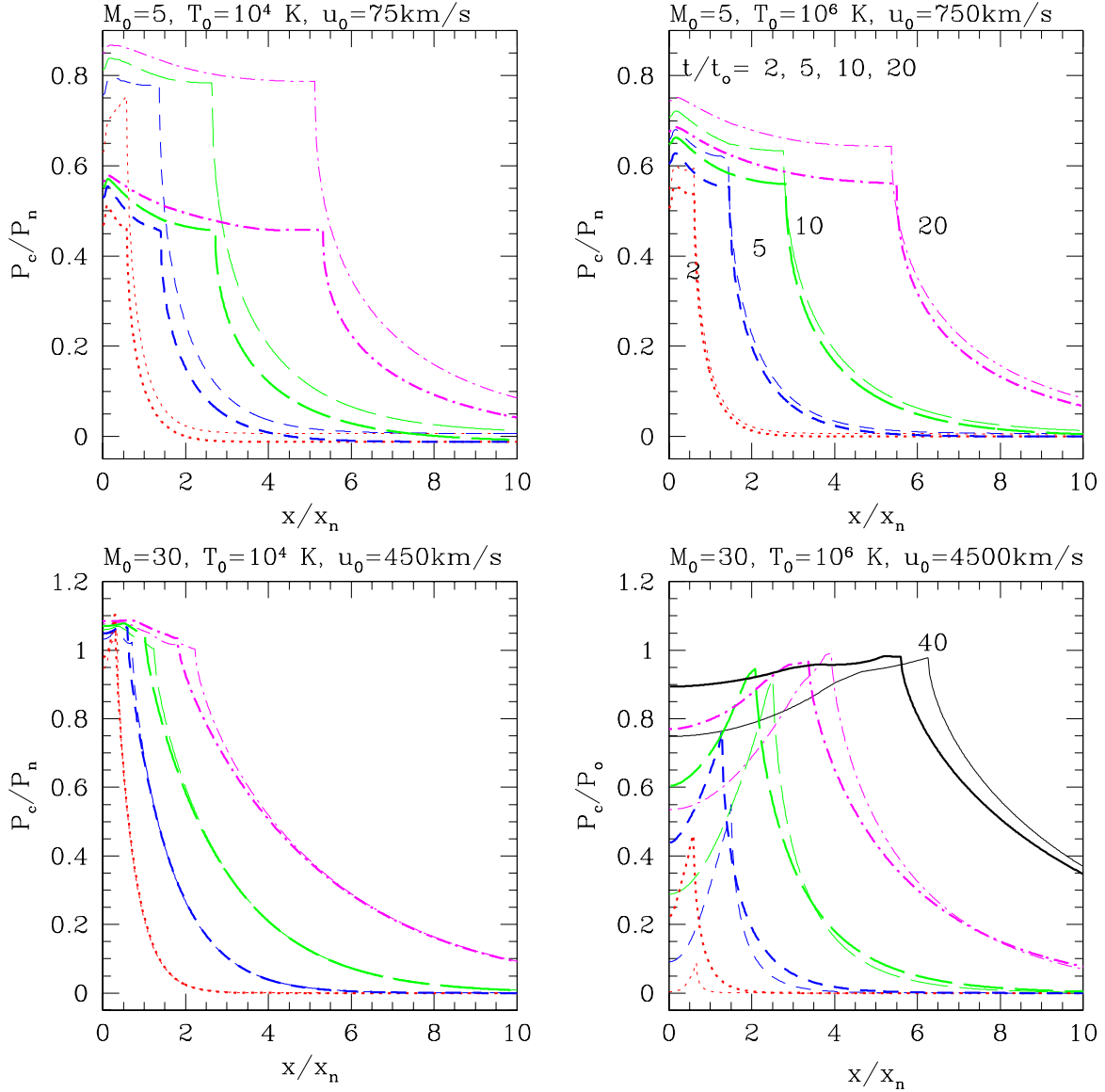}
\figcaption{
Time evolution of the CR pressure for the following models:
1) $M_0=5$ and $T_0=10^4$ K (upper left panel),
2) $M_0=5$ and $T_0=10^6$ K (upper right panel),
3) $M_0=30$ and $T_0=10^4$ K (lower left panel), and
4) $M_0=30$ and $T_0=10^6$ K (lower right panel).
Heavy lines are for the models without pre-existing CRs,
while light lines are for the models with $P_{c,0}=0.25P_{g,0}$.
\label{fig3}}
\end{figure}
\clearpage

\begin{figure}
\plotone{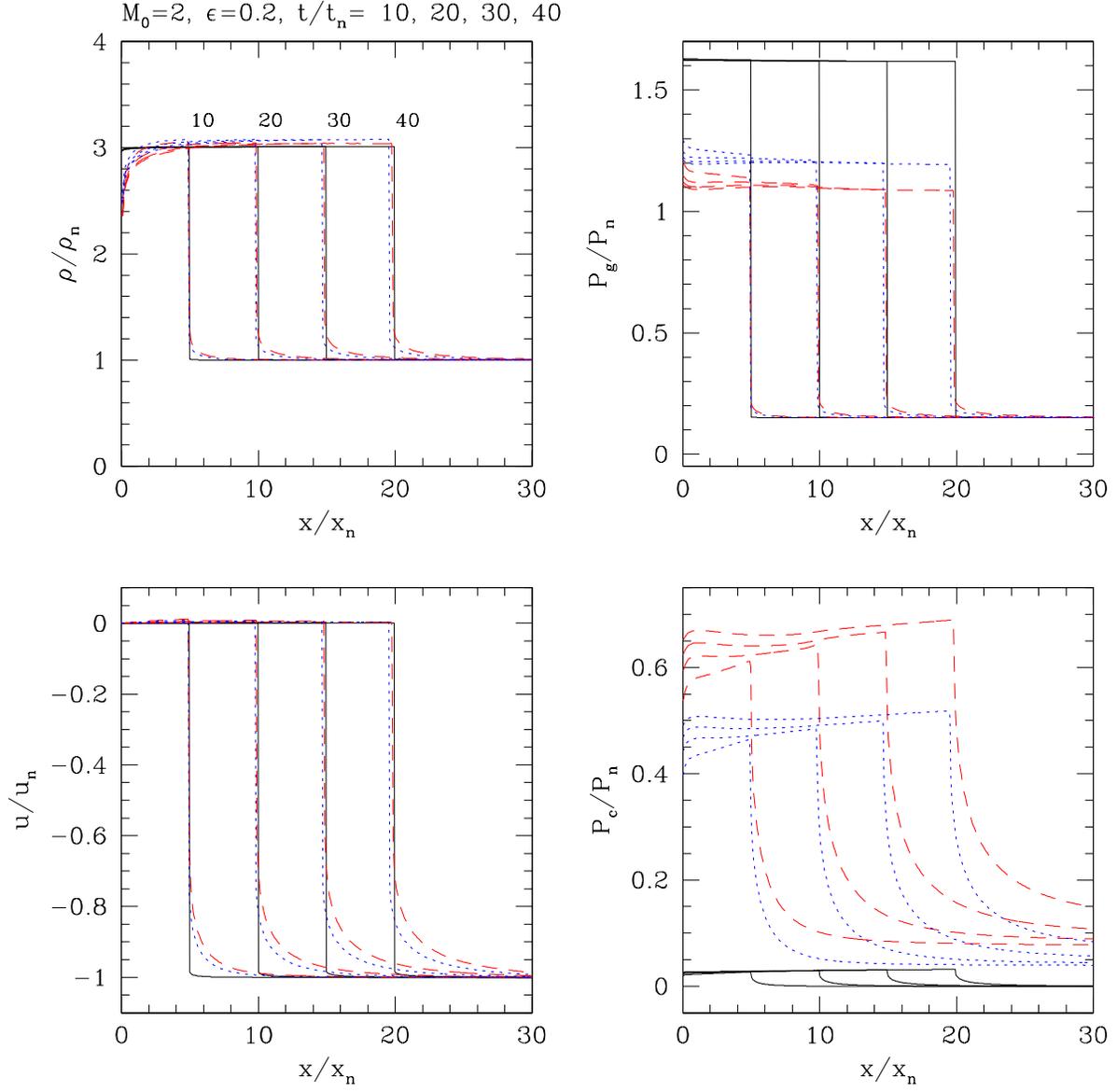}
\figcaption{
Same as Figure 1 except $M_0=2$, $T_0=10^6$K and  $u_{\rm n}=300~\kms$. 
The solid lines are for the model with $P_{c,0}=0.0$,
the dotted lines for the model with $P_{c,0}=0.25 P_{g,0}$, 
and the dashed lines for the model with $P_{c,0}=0.5 P_{g,0}$. 
\label{fig4}}
\end{figure}
\clearpage

\begin{figure}
\plotone{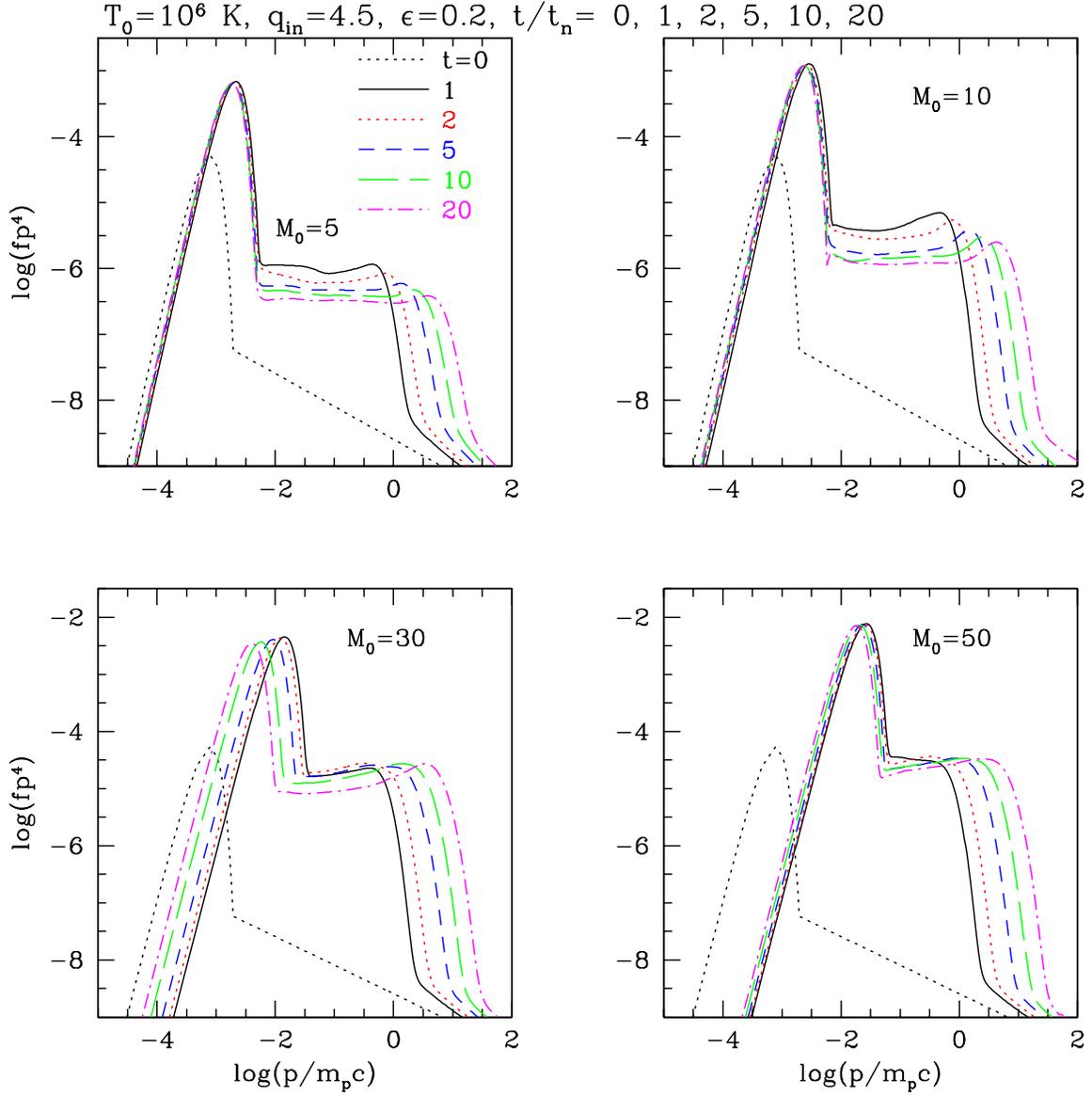}
\figcaption{
Evolution of the CR distribution function at the shock, represented
as $g=p^4f(p)$, is plotted for
the models of $M_0=$ 5, 10, 30, and 50 with $T_0=10^6$ K and
$f_{up} \propto (p/p_M)^{-4.5}$.
The CR spectrum of the preshock flow is represented by the dotted line. 
For all models shown here the upstream CR pressure is $P_{c,0}/P_{g,0}=0.25$.
\label{fig5}}
\end{figure}
\clearpage

\begin{figure}
\plotone{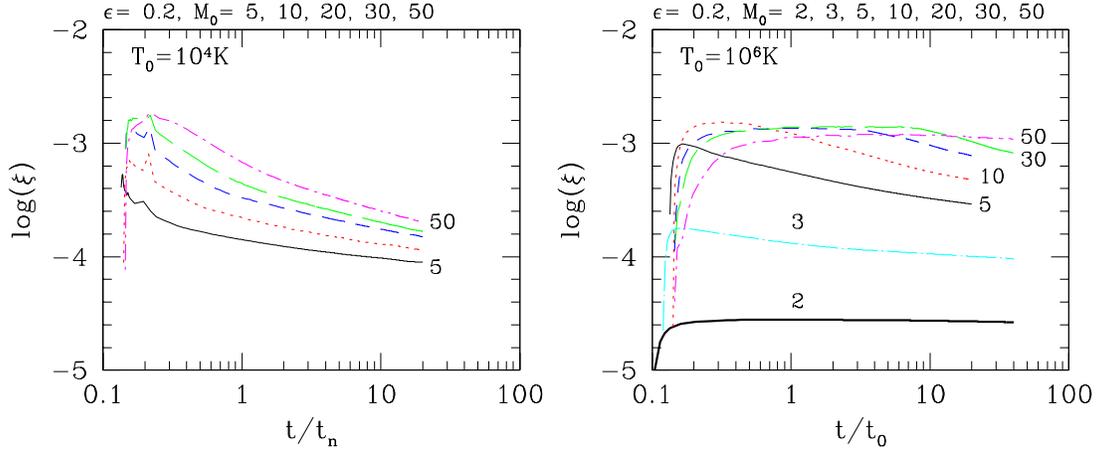}
\figcaption{
Time averaged injection efficiency, $\xi(t)$, for models without pre-existing CRs.
Left panel shows the models with $M_0=5-50$ and $T_0=10^4$ K,
while right panel shows the models with $M_0=2-50$ and $T_0=10^6$ K.
\label{fig6}}
\end{figure}
\clearpage

\begin{figure}
\plotone{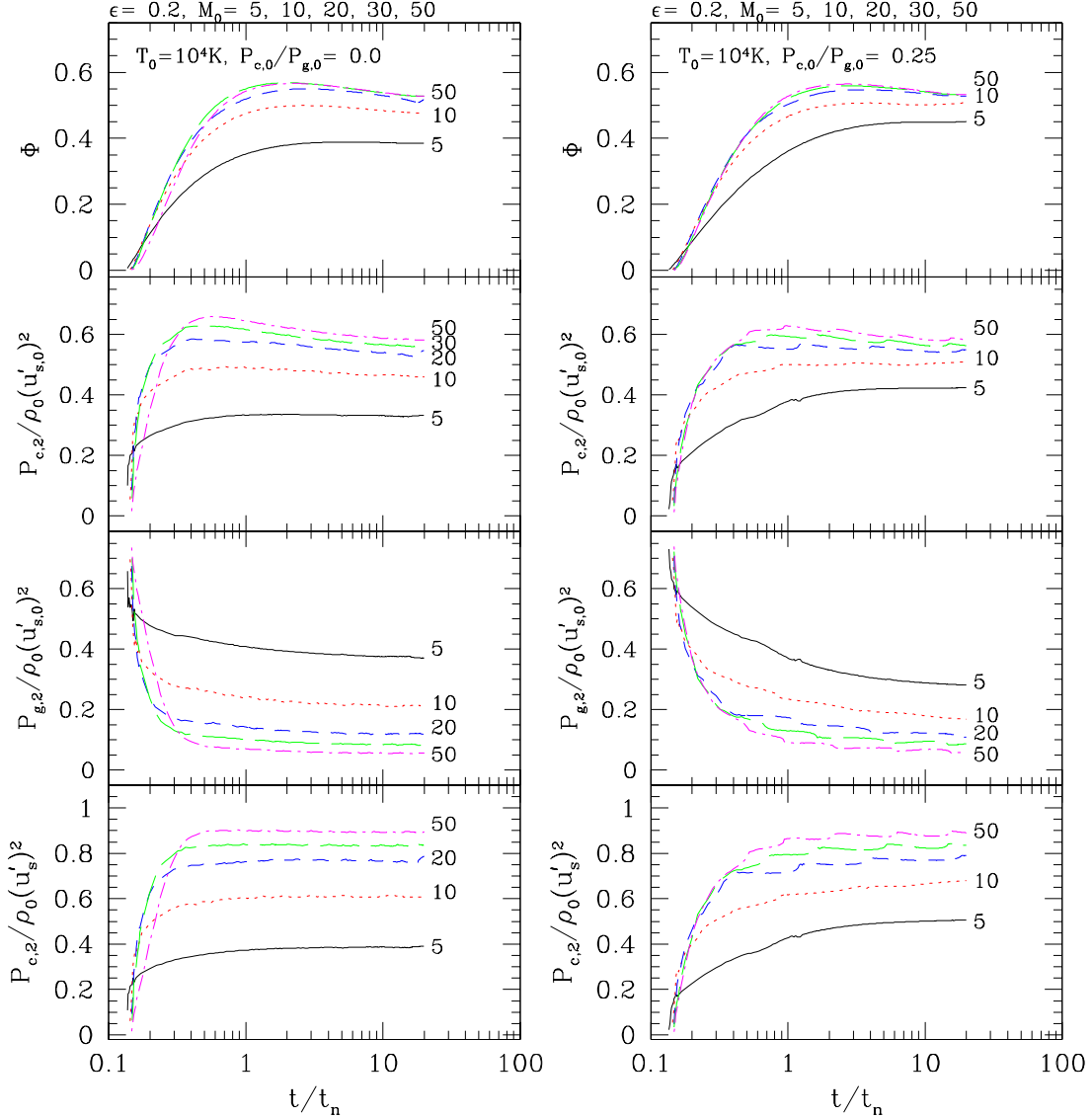}
\figcaption{
The ratio of total CR energy in the simulation box 
to the kinetic energy in the initial shock rest frame
that has entered the simulation box from upstream, $\Phi(t)$, 
the postshock CR pressure, $P_{c,2}$, and gas pressure, $P_{g,2}$,
in units of upstream ram pressure in the {\it initial} shock frame, 
$\rho_0 u_{s,0}^{\prime~2}$
and the postshock CR pressure in units of upstream ram pressure 
in the {\it instantaneous} shock frame, $\rho_0 u_{s}^{\prime~2}$
Left panels show the models with $M_0=5-50$ and $T_0=10^4$ K but
without pre-existing CRs.
Right panels for the same models but with the upstream CR pressure of
$P_{c,0}=0.25 P_{g,0}$ and $f_{up} \propto (p/p_M)^{-4.7}$. 
\label{fig7}}
\end{figure}
\clearpage

\begin{figure}
\plotone{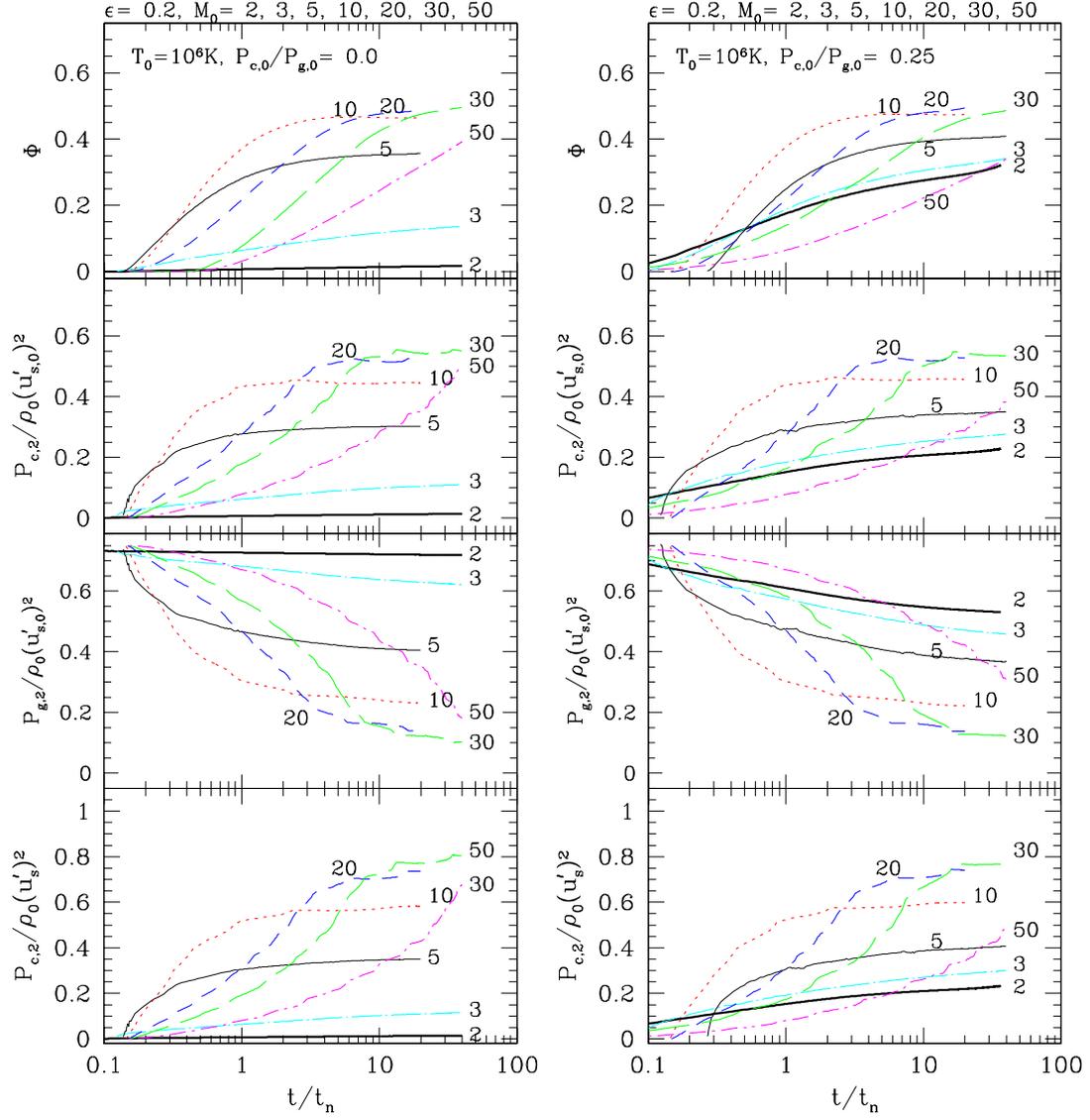}
\figcaption{
Same as figure 9 except that
the models with $T_0=10^6$ K are shown.
Left panels for the models without pre-existing CRs, while
right panels for the models with $P_{c,0}=0.25P_{g,0}$ and
$f_{up} \propto (p/p_M)^{-4.5}$ .
\label{fig8}}
\end{figure}
\clearpage

\begin{figure}
\plotone{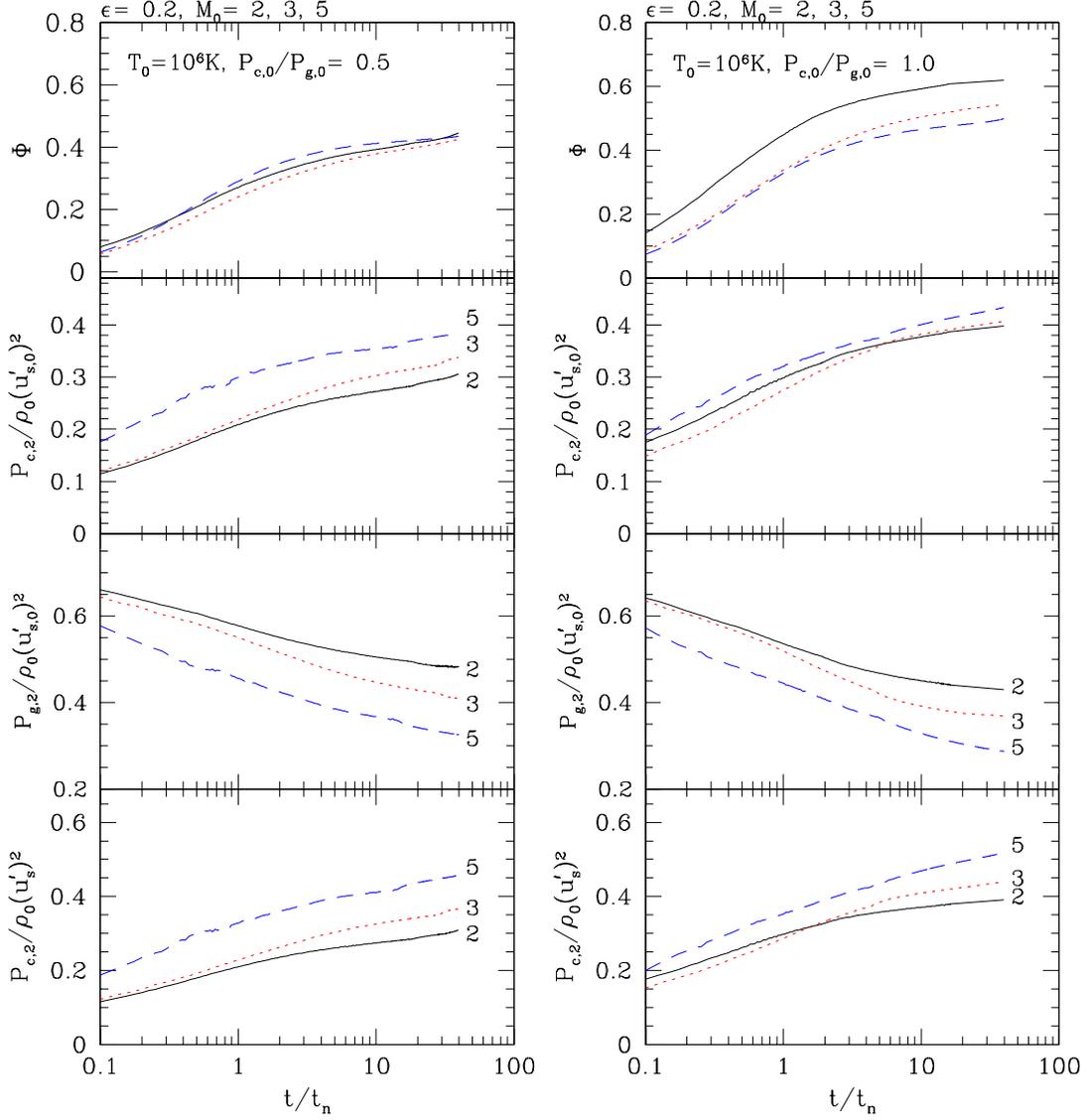}
\figcaption{
Same as figure 9 except that 
the low Mach models with $T_0=10^6$ K are shown.
Left panels show the models with pre-existing CRs of 
$P_{c,0}=0.5P_{g,0}$ and $f_{up} \propto (p/p_M)^{-4.4}$,
while right panels for the models with  
$P_{c,0}=P_{g,0}$ and $f_{up} \propto (p/p_M)^{-4.3}$.
Solid line is for $M_0=2$ model, dotted line for $M_0=3$ model,
and dashed line for $M_0=5$ models. 
\label{fig9}}
\end{figure}
\clearpage

\begin{figure}
\plotone{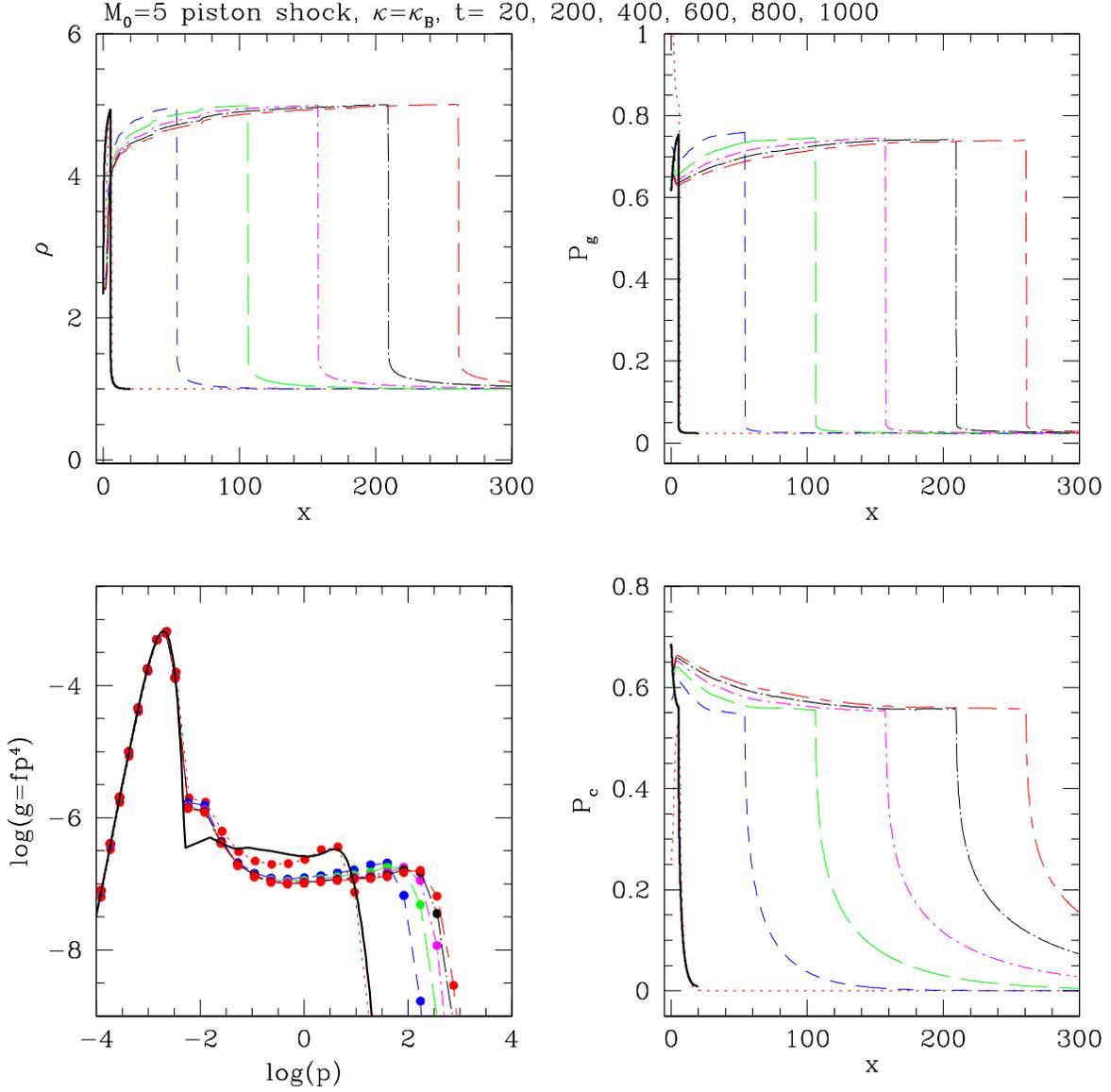}
\figcaption{
Time evolution of the $M_0=5$ model without pre-existing CRs
calculated with "Coarse-Grained Momentum Volume" method is shown up to 
$t/t_{\rm n}=$1000.
The leftmost profile corresponds to the earliest time, $t/t_{\rm n}=$ 20. 
For comparison, the results of "Fine-Grained Momentum Volume" simulation 
are shown at the same time by the heavy solid line. 
\label{fig10}}
\end{figure}
\clearpage

\begin{figure}
\plotone{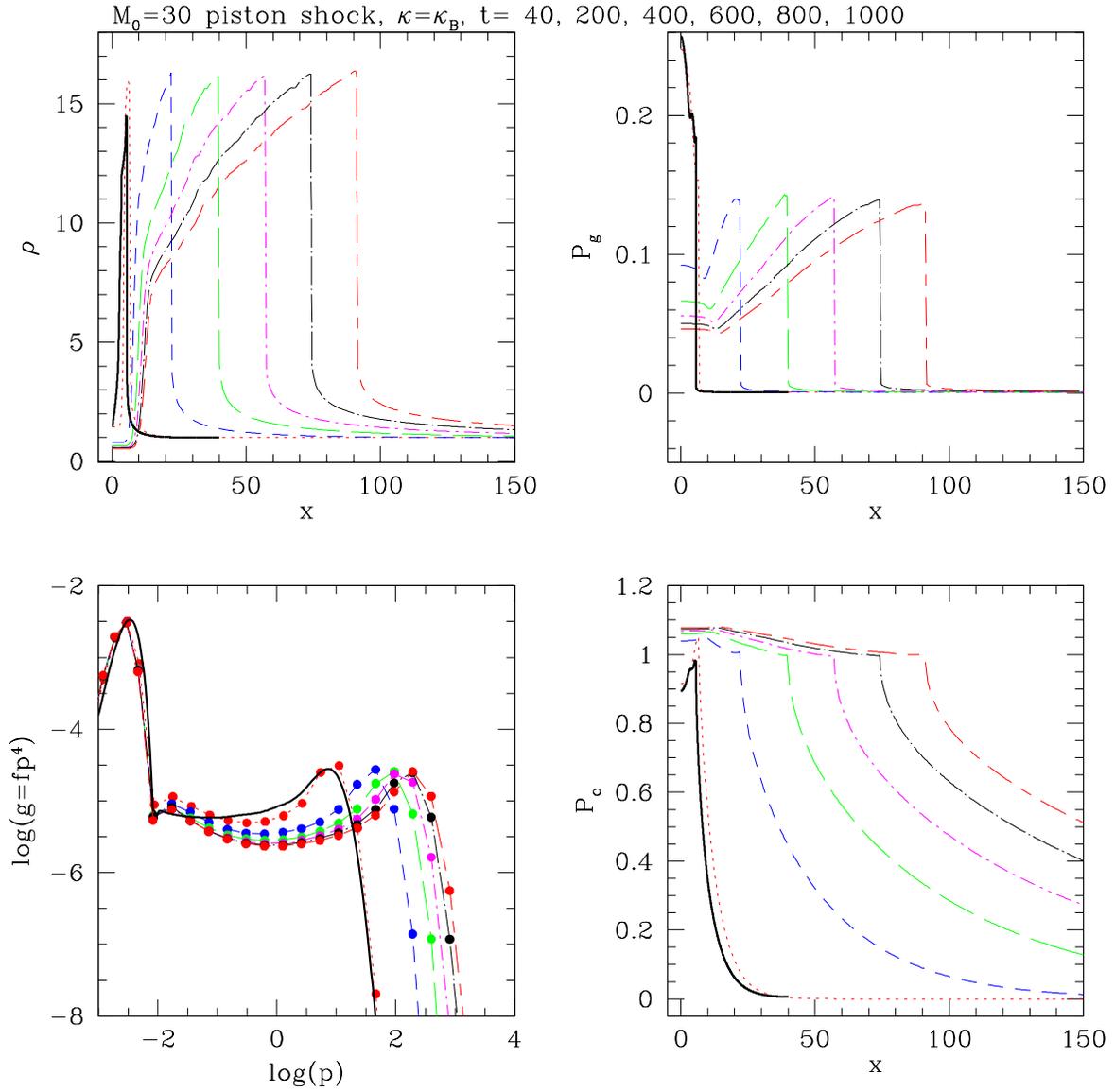}
\figcaption{
Same as Figure 10 except $M_0=30$.
The results from two simulations are compared at $t/t_{\rm n}=$ 40.
\label{fig11}}
\end{figure}
\clearpage

\begin{figure}
\plotone{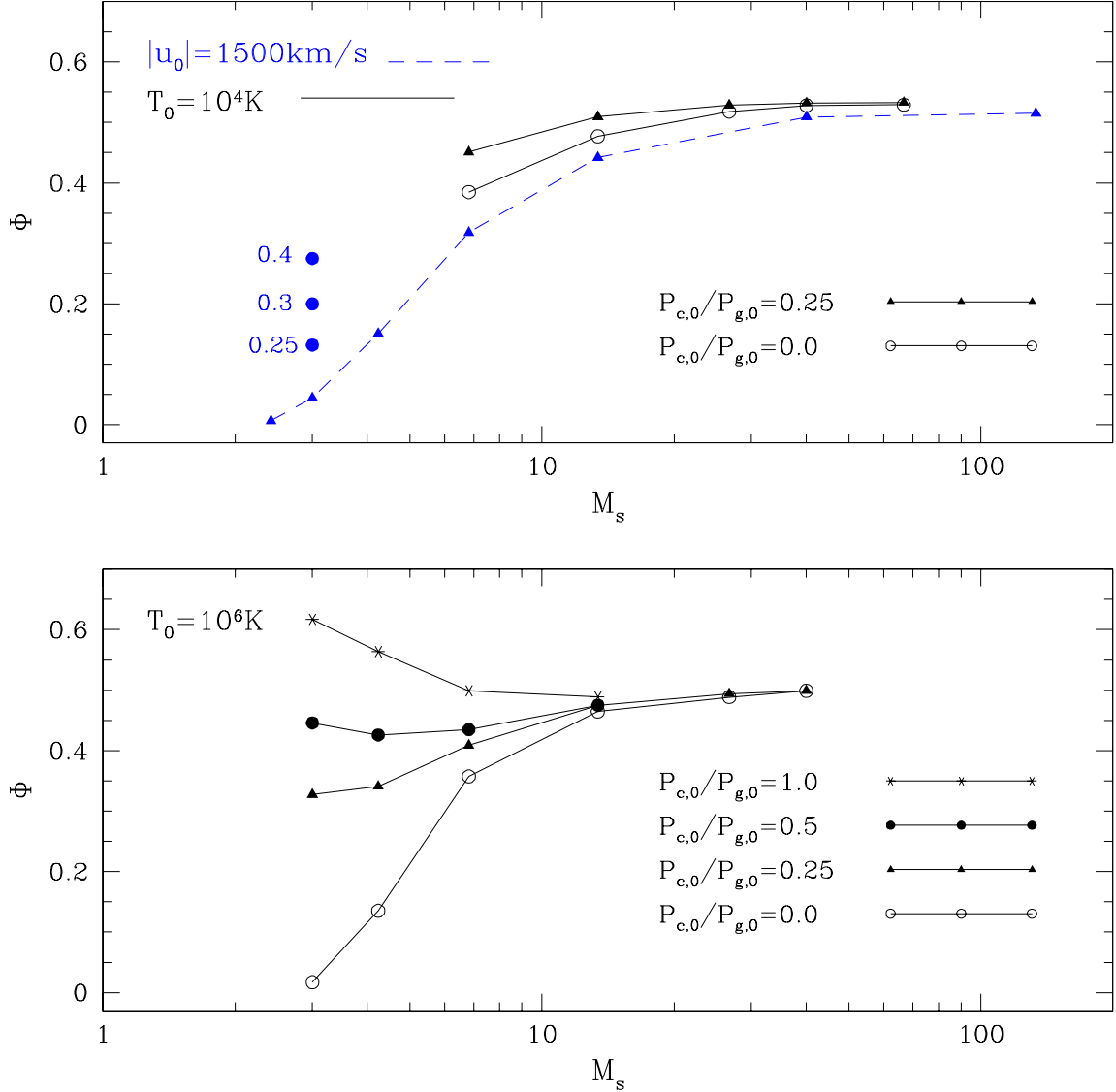}
\figcaption{
The CR energy ratio, $\Phi$, at the simulation termination time of our
simulations as a function of the shock Mach number, $M_s$.
{\it Upper panel}: 
models with $T_0=10^4$ K, $u_s=(15\kms)M_s$, $\epsilon=0.2$, and no pre-existing CRs
(solid line with open circles),
models with $T_0=10^4$ K, $u_s=(15\kms)M_s$, $\epsilon=0.2$, and $P_{c,0}=0.25P_{g,0}$ 
(solid line with filled triangles).
For comparison we also show ``constant $u_0$ models with 
$|u_0|=1500 \kms$, $T_0=10^4{\rm K}(100/M_0)$, $\epsilon=0.2$, and 
no pre-existing CRs in the upper panel (dashed line with filled triangles).
Three filled circles are labeled with the value of $\epsilon=$ 0.25, 0.3, 0.4
and for the models with $M_s=3$, $T_0=5\times 10^5$K and no pre-existing CRs.
{\it Lower panel}: 
models with $T_0=10^6$ K, $u_s=(150\kms)M_s$, $\epsilon=0.2$, and 
a various level of pre-existing CR pressure.
See Table 1 for the relation between the accretion Mach number $M_0$ and
$M_s$.
\label{fig12}}
\end{figure}
\clearpage

\begin{deluxetable} {cccccc}
\tablecaption{Model Parameters for Flows Upstream of Shocks}
\tablehead{
\colhead {$T_0$} & \colhead{$|u_0|$} & \colhead{$M_0^{\rm a}$} & \colhead{$M_s^{\rm b}$}&
\colhead{$P_{c,0}/P_{g,0}$ }& \colhead{$q_{in}~^{\rm c}$} \\
\colhead {(K)} &\colhead {($\kms$)}
}
\startdata
$10^4$  & $(15)M_0$ & 5,~10,~20,~30,~50 &6.8,~13.4,~26.7,~40.,~66.7& 0.0& ... \\
$10^4$  & $(15)M_0$ & 5,~10,~20,~30,~50&6.8,~13.4,~26.7,~40.,~66.7 & 0.25& 4.7 \\
$10^6$  & $(150)M_0$ & 2,~3,~5,~10,~20,~30,~50 &3.,~4.2,~6.8,~13.4,~26.7,~40.,~66.7& 0.0& ... \\
$10^6$  & $(150)M_0$ & 2,~3,~5,~10,~20,~30,~50&3.,~4.2,~6.8,~13.4,~26.7,~40.,~66.7 & 0.25& 4.5 \\
$10^6$  & $(150)M_0$ & 2,~3,~5,~10&3.,~4.2,~6.8,~13.4 & 0.5& 4.4 \\
$10^6$  & $(150)M_0$ & 2,~3,~5,~10&3.,~4.2,~6.8,~13.4 & 1.0& 4.3 \\
$10^8M_0^{-2}$  & $ 1500 $ & 1.5,~2,~3,~5,~10,~30,~100 & 2.4,~3.,~4.2,~6.8,~13.4,~40.,~133.& 0.0& ... \\
\enddata

\tablenotetext{a} {Mach number of piston with respect to inflow.}
\tablenotetext{b} {Mach number of initial gas shock with respect to infall.}
\tablenotetext{c} {Power-law index of upstream CR population, \ie
$f_{up} \propto p^{-q_{in}}$.}

\end{deluxetable}

\end{document}